\documentclass[journal, 12pt, draftclsnofoot, onecolumn]{IEEEtran}

\usepackage{acronym}
\acrodef{AOA}{angle-of-arrival}
\acrodef{AOD}{angle-of-departure}
\acrodef{MISO}{multiple-input single-output}

\usepackage{amsmath,amssymb,epsfig,psfrag,cite,subfigure}
\include{macros}
\usepackage{graphicx}
\usepackage{epstopdf}

\usepackage{bm}

\usepackage{tcolorbox}

\usepackage{soul}

\usepackage{amsthm}

\usepackage{nicefrac}

\usepackage{lipsum,multicol}

\usepackage{algorithm,algorithmic}

\usepackage{enumitem}
\usepackage{mwe}    
\usepackage{epsfig,psfrag}
\usepackage{subfigure}
\usepackage{color}
\usepackage{url}
\usepackage{mathtools,xparse}

\usepackage{gensymb}

\usepackage[multiple]{footmisc}

\usepackage{accents}

\makeatletter
\newcommand*\rel@kern[1]{\kern#1\dimexpr\macc@kerna}
\newcommand*\widebar[1]{%
  \begingroup
  \def\mathaccent##1##2{%
    \rel@kern{0.8}%
    \overline{\rel@kern{-0.8}\macc@nucleus\rel@kern{0.2}}%
    \rel@kern{-0.2}%
  }%
  \macc@depth\@ne
  \let\math@bgroup\@empty \let\math@egroup\macc@set@skewchar
  \mathsurround\z@ \frozen@everymath{\mathgroup\macc@group\relax}%
  \macc@set@skewchar\relax
  \let\mathaccentV\macc@nested@a
  \macc@nested@a\relax111{#1}%
  \endgroup
}
\makeatother

\DeclareFontFamily{U}{mathx}{\hyphenchar\font45}
\DeclareFontShape{U}{mathx}{m}{n}{
	<5> <6> <7> <8> <9> <10>
	<10.95> <12> <14.4> <17.28> <20.74> <24.88>
	mathx10
}{}
\DeclareSymbolFont{mathx}{U}{mathx}{m}{n}
\DeclareFontSubstitution{U}{mathx}{m}{n}

\allowdisplaybreaks

\theoremstyle{remark}

\newtheoremstyle{mytheoremstyle} 
    {\topsep}                    
    {\topsep}                    
    {\upshape}                   
    {.5em}                           
    {\itshape}                   
    {.}                          
    {.5em}                       
    {}  

\theoremstyle{plain}

\newtheoremstyle{iremark}
  {\topsep}   
  {\topsep}   
  {\upshape}  
  {0.2in}       
  {\itshape}  
  {.}         
  {5pt plus 1pt minus 1pt} 
  {\thmname{#1}\thmnumber{ \itshape#2}\thmnote{ (#3)}} 

\newtheorem{theorem}{Theorem}
\newtheorem{lemma}[theorem]{Lemma}

\theoremstyle{definition}
\newtheorem*{proof}{Proof}

\theoremstyle{definition}

\DeclarePairedDelimiter\abs{\lvert}{\rvert}%
\DeclarePairedDelimiter\absbig{\Big\lvert}{\Big\rvert}%

\makeatletter \renewcommand\d[1]{\ensuremath{%
		\;\mathrm{d}#1\@ifnextchar\d{\!}{}}}
\makeatother

\newcommand{\norm}[1]{\left\lVert#1\right\rVert}

\newcommand{\thn}[1]{ {#1^{\rm{th} } } }

\newcommand{\Eee}{\mathbb{E}}

\newcommand{\Tcp}{ T_{\rm{cp}} }

\newcommand{\FF}{ \mathbf{F} }

\newcommand{\deltaf}{ \Delta f }
\newcommand{\fc}{ f_c }
\newcommand{\Ntx}{ N_{\rm{T}} }
\newcommand{\Nrx}{ N_{\rm{R}} }

\newcommand{\atx}{ \aaa_{\rm{T}} }
\newcommand{\arx}{ \aaa_{\rm{R}} }

\newcommand{\ssscp}{\sss_{\rm{CP}}}
\newcommand{\scpi}{s_{{\rm{CP}},i}}
\newcommand{\scpig}[1]{s_{{\rm{CP}},#1}}

\newcommand{\yycp}{\yy_{\rm{CP}}}

\newcommand{\Ktilde}{\widetilde{K}}

\newcommand{\hhcom}{\hh_{ {\rm{com}} }}
\newcommand{\Kric}{ J_{\rm{R}} }

\newcommand{\gtx}{g_{\rm{tx}}}
\newcommand{\grx}{g_{\rm{rx}}}
\newcommand{\gtxmat}{\mathbf{G}_{\rm{tx}}}
\newcommand{\grxmat}{\mathbf{G}_{\rm{rx}}}

\newcommand{\etatilde}{\widetilde{\eta}}
\newcommand{\etabar}{\widebar{\eta}}

\newcommand{\kbar}{ \widebar{K} }

\newcommand{\pfa}{P_{\rm{fa}}}

\newcommand{\snr}{{\rm{SNR}}}
\newcommand{\snrrad}{{\rm{SNR}}_{\rm{rad}}}

\newcommand{\Rmax}{ R_{\rm{max}} }
\newcommand{\vmax}{ v_{\rm{max}} }

\newcommand{\alphahat}{ \widehat{\alpha} }

\newcommand{\taumaxisi}{ \tau^{\rm{ISI}}_{\rm{max}} }
\newcommand{\rmaxisi}{ R^{\rm{ISI}}_{\rm{max}} }
\newcommand{\numaxici}{ \nu^{\rm{ICI}}_{\rm{max}} }
\newcommand{\vmaxici}{ v^{\rm{ICI}}_{\rm{max}} }

\newcommand{\rlmmmse}{ \mathbf{R}_{\rm{LMMSE}} }

\newcommand{\taumax}{ \tau_{\rm{max}} }

\newcommand{\numax}{ \nu_{\rm{max}} }

\newcommand{\taut}{  \widetilde{\tau} }
\newcommand{\thetat}{  \widetilde{\theta} }

\newcommand{\nut}{  \widetilde{\nu} }

\newcommand{\tauk}{ \tau_k }

\newcommand{\vk}{ v_k }
\newcommand{\nuk}{ \nu_k }

\newcommand{\mtCN}{{\mathcal{CN}}}

\newcommand{\traceee}{ {\rm{tr}}  }

\newcommand{\tracebig}[1]{ {{{\rm{tr}}\Big( #1 \Big)}}  }
\newcommand{\tracesmall}[1]{ {{{\rm{tr}}\left( #1 \right)}}  }

\newcommand{\vecc}[1]{ {\rm{vec}}\hspace{-0.02in}\left(#1\right)  }

\newcommand{\veccinv}[1]{ {\rm{reshape}}_{N,M}\left(#1\right)  }

\newcommand{\diag}[1]{ {\rm{diag}}\left(#1\right)  }

\newcommand{\diagb}[1]{ {\rm{diag}}\big(#1\big)  }

\newcommand{\Imatrix}{{ \boldsymbol{\mathrm{I}} }}

\newcommand{\Rs}{{ \mathcal{R} }}
\newcommand{\Rtau}{{ \mathcal{R}_{\tau} }}
\newcommand{\Rnu}{{ \mathcal{R}_{\nu} }}
\newcommand{\Rtheta}{{ \mathcal{R}_{\theta} }}

\newcommand{\aaa}{\mathbf{a}}
\newcommand{\cc}{ \mathbf{c} }

\newcommand{\bb}{ \mathbf{b} }

\newcommand{\bbisi}{ \bb^{{\rm{ISI}}} }
\newcommand{\ccici}{ \cc^{{\rm{ICI}}} }

\newcommand{\nuhat}{{ \widehat{\nu} }}
\newcommand{\tauhat}{{ \widehat{\tau} }}
\newcommand{\thetahat}{{ \widehat{\theta} }}

\newcommand{\boldzero}{{ {\boldsymbol{0}} }}
\newcommand{\boldone}{{ {\boldsymbol{1}} }}

\newcommand{\boldS}{ \mathbf{S} }

\newcommand{\boldY}{ \mathbf{Y} }
\newcommand{\boldYs}{ \boldY^{{\rm{single}}} }

\newcommand{\boldX}{ \mathbf{X} }
\newcommand{\boldXdd}{ \mathbf{X}^{\rm{DD}} }

\newcommand{\boldYcomdd}{ \boldY_{\rm{com}}^{\rm{DD}} }

\newcommand{\boldW}{ \mathbf{W} }

\newcommand{\boldZ}{ \mathbf{Z} }

\newcommand{\ycomcp}{ y_{\rm{com},\rm{CP}} }

\newcommand{\boldQ}{ \mathbf{Q} }
\newcommand{\boldH}{ \mathbf{H} }

\newcommand{\boldG}{ \mathbf{G} }

\newcommand{\boldC}{ \mathbf{C} }
\newcommand{\boldB}{ \mathbf{B} }

\newcommand{\boldD}{ \mathbf{D} }

\newcommand{\boldA}{ \mathbf{A} }

\newcommand{\boldLambda}{ \mathbf{\Lambda} }

\newcommand{\qq}{ \mathbf{q} }
\newcommand{\yy}{ \mathbf{y} }
\newcommand{\pp}{ \mathbf{p} }
\newcommand{\xx}{ \mathbf{x} }
\newcommand{\hh}{ \mathbf{h} }
\newcommand{\ww}{ \mathbf{w} }

\newcommand{\zztilde}{ \widetilde{\zz} }

\newcommand{\ztilde}{ \widetilde{z} }
\newcommand{\zz}{ \mathbf{z} }

\newcommand{\ppbar}{ \widebar{\pp} }

\newcommand{\xxddbar}{ \widebar{\xx}^{\rm{DD}} }

\newcommand{\xxhat}{ \widehat{\xx} }

\newcommand{\ggb}{ \mathbf{g} }

\newcommand{\Lambdalog}{ \Lambda^{{\rm{log}}} }

\newcommand{\deltatau}{\Delta_{\tau}}

\newcommand{\sss}{ \mathbf{s} }

\newcommand{\boldYcom}{ \boldY_{\rm{com}} }

\newcommand{\boldHdd}{ \boldH_{\rm{DD}} }
\newcommand{\boldHt}{ \boldH_{\rm{T}} }

\newcommand{\yycom}{ \yy_{\rm{com}} }
\newcommand{\yycomdd}{ \yy_{\rm{com}}^{\rm{DD}} }
\newcommand{\xxdd}{ \xx^{\rm{DD}} }
\newcommand{\xxddhat}{ \xxhat^{\rm{DD}} }

\newcommand{\DDrad}{ \boldD_{\rm{rad}} }
\newcommand{\DDcom}{ \boldD_{\rm{com}} }

\newcommand{\boldUps}{ \mathbf{\Upsilon} }

\newcommand{\boldGamma}{ \mathbf{\Gamma} }

\newcommand{\transpose}[1]{ {#1}^{T} }

\newcommand{\complexset}[2]{ \mathbb{C}^{#1 \times #2}  }
\newcommand{\complexsett}{ \mathbb{C}  }

\newcommand{\realset}[2]{ \mathbb{R}^{#1 \times #2}  }

\newcommand{\alphat}{ \widetilde{\alpha} }

\newcommand{\conj}[1]{ {#1}^{\ast} }

\newcommand{\betab}{ {\boldsymbol{\beta}} }

\newcommand{\betabopt}{ \betab^{\rm{opt}} }
\newcommand{\ppopt}{ \pp^{\rm{opt}} }
\newcommand{\qqopt}{ \qq^{\rm{opt}} }

\newcommand{\lambdab}{ {\boldsymbol{\lambda}} }

\graphicspath{{./Figures/}}

\begin{document}
\bstctlcite{IEEEexample:BSTcontrol}
\title{Integrated Sensing and Communications with MIMO-OTFS}

\author{Musa Furkan Keskin, \textit{Member, IEEE}, Carina Marcus, Olof Eriksson, Alex Alvarado, \textit{Senior Member, IEEE}, Joerg Widmer, \textit{Fellow, IEEE}, and Henk Wymeersch, \textit{Senior Member, IEEE}\thanks{Musa Furkan Keskin and Henk Wymeersch are with the Department of Electrical Engineering, Chalmers University of Technology, SE 41296 Gothenburg, Sweden (e-mail: furkan@chalmers.se). Carina Marcus and Olof Eriksson are with Veoneer Sweden AB, SE 44737 Vårgårda, Sweden. Alex Alvarado is with the Department of Electrical Engineering, Eindhoven University of Technology, 5600 MB Eindhoven, The Netherlands. Joerg Widmer is with IMDEA Networks, 28918 Madrid, Spain. This work is supported, in part, by Vinnova grant 2021-02568, MSCA-IF grant 888913 (OTFS-RADCOM), 
and the Dutch Technology Foundation TTW, part of the Netherlands Organisation for Scientific Research (NWO), which is partly funded by the Ministry of Economic Affairs under the project Integrated Cooperative Automated Vehicles (i-CAVE).}}

\maketitle

\begin{abstract}
    Orthogonal time frequency space (OTFS) is a promising alternative to orthogonal frequency division multiplexing (OFDM) for high-mobility  communications. We propose a novel multiple-input multiple-output (MIMO) integrated sensing and communication (ISAC) system based on OTFS modulation. We begin by deriving new sensing and communication signal models for the proposed MIMO-OTFS ISAC system that explicitly capture inter-symbol interference (ISI) and inter-carrier interference (ICI) effects. We then develop a generalized likelihood ratio test (GLRT) based multi-target detection and delay-Doppler-angle estimation algorithm for MIMO-OTFS radar sensing that can simultaneously mitigate and exploit ISI/ICI effects, to prevent target masking and surpass standard unambiguous detection limits in range/velocity. Moreover, considering two operational modes (search/track), we propose an adaptive MIMO-OTFS ISAC transmission strategy. For the search mode, we introduce the concept of delay-Doppler (DD) multiplexing, 
    enabling omnidirectional probing of the environment and large virtual array at the OTFS radar receiver. For the track mode, we pursue a directional transmission approach and  design an OTFS ISAC optimization algorithm in spatial and DD domains, seeking the optimal trade-off between radar signal-to-noise ratio (SNR) and achievable rate. Simulation results verify the effectiveness of the proposed sensing algorithm and reveal valuable insights into OTFS ISAC trade-offs under varying communication channel characteristics.

\end{abstract}

\section{Introduction}

\subsection{Background and Motivation}
As 5G systems are being rolled out, the time has come to conceive and develop 6G communication systems. There are now several initiatives in Europe, the USA, and Asia to define what 6G will be in terms of use cases and requirements \cite{6g_vision_2023,6g_hexax,6g_wp3_hexax}. As with all previous generations, one requirement will be a 10-fold increase in peak data rate. Unlike previous generations, there is now greater emphasis placed on integrated sensing and communications (ISAC) \cite{Fan_ISAC_6G_JSAC_2022}, driven not only by localization/sensing use cases but also the inherent geometric nature of the wireless propagation channel \cite{Lima6Gsensing20,wymeersch2020radio}. 

In pursuit of higher data rates, 
lower latency, and higher sensing accuracies, we have no choice but to consider larger carrier frequencies, above the 24 GHz band in 5G, as this is where larger bandwidths are available \cite{5g_6g_isac_2021}. At lower frequencies, (despite intense competition) OFDM has remained the communication waveform of choice, due to its robustness to multipath, simple equalization, straightforward integration with multi-antenna systems, and high flexibility in terms of power and rate allocation \cite{banelli2014modulation}. In addition, OFDM proves to be suitable for ISAC with standard FFT-processing, in both mono-static and bi-static configurations \cite{ofdm_radar_correlation_TAES_2020,Fan_ISAC_6G_JSAC_2022}. However, at 6G frequencies, OFDM is challenged by several effects, which necessitates the consideration of alternative modulation formats. Firstly, OFDM suffers from a high peak-to-average power ratio (PAPR), leading to reduced  power efficiency, which becomes a limiting factor at high carriers. Secondly, OFDM requires frequent adaptation due to mobility and fading, which would lead to prohibitive overheads at high carriers due to the short coherence times \cite{banelli2014modulation}. Moreover, the robustness to multipath comes at a cost of inserting a cyclic prefix (CP) between OFDM symbols, resulting in a rate loss, to combat inter-symbol interference (ISI). Finally, for radar, OFDM is sensitive to inter-carrier-interference (ICI), resulting from Doppler shifts under high target velocities \cite{MIMO_OFDM_ICI_JSTSP_2021}. 

These drawbacks of OFDM have sparked renewed interest in alternative modulation schemes, in particular with favorable properties in terms of ISAC performance. Orthogonal time frequency space (OTFS) has become a promising candidate in this respect, as it has lower PAPR \cite{papr_otfs_2022}, requires less frequent adaptation \cite{hadani2017orthogonal}, incurs a much lower CP overhead and can cope with much larger Doppler shifts \cite{otfs_ofdm_comp_TWC_2022}. As opposed to frequency-time (FT) domain multiplexing in OFDM modulation, OTFS multiplexes the data symbols in the delay-Doppler (DD) domain. This implies that a time-varying channel with constant Doppler will appear time-invariant to OTFS \cite{OTFS_SBL_TWC_2022}, which can be exploited to improve bit error rate (BER) performance in high-mobility scenarios \cite{OTFS_CE_TSP_2019,isac_otfs_jstsp_2021}. From a practical viewpoint, efficient Zak transform-based implementations of OTFS have been recently proposed \cite{OTFS_mag_2022,lampel2022orthogonal}. In addition to its advantages for communications, OTFS can also bring potential benefits for radar sensing since radar detections are of the form of range (delay) and velocity (Doppler) tuples \cite{ISAC_OTFS_JSAC_2022}. Overall, OTFS stands out as a natural candidate for ISAC, especially in high-mobility vehicular applications, as evidenced by recent activity in this area \cite{Gaudio_MIMO_OTFS_Hybrid,otfs_radar_2019,OTFS_RadCom_TWC_2020,MIMO_OTFS_Radar_2020,OTFS_IOT_2021,beamspaceMIMO_OTFS_2022,ISAC_OTFS_JSAC_2022,OTFS_ISAC_part3_2022,ofdm_otfs_comparison_2022,isac_otfs_jstsp_2021}. 

\subsection{Related Work on OTFS ISAC}
As a common approach in the OTFS ISAC literature, target detection and range-velocity estimation are performed by converting the received time-domain signal back to the DD domain \cite{Gaudio_MIMO_OTFS_Hybrid,otfs_radar_2019,OTFS_RadCom_TWC_2020,beamspaceMIMO_OTFS_2022}. This leads to high complexity, e.g., requiring iterative interference cancellation-based processing due to significant side-lobe levels\footnote{OTFS can lead to signal processing challenges also at the communications side, especially related to equalization due to large pilot overhead and high signal-to-noise ratio (SNR) requirement to obtain accurate channel state information \cite{OTFS_Eq_Learning_TWC_2022}.}\cite{Gaudio_MIMO_OTFS_Hybrid}. In addition, the expression of the radar signal in the DD domain is quite complicated (see, e.g., \cite[Eq.~(12)]{OTFS_RadCom_TWC_2020}, \cite[Eq.~(11)]{Gaudio_MIMO_OTFS_Hybrid}, \cite[Eq.~(11)]{beamspaceMIMO_OTFS_2022}), making it difficult to derive insights into the structure of the OTFS signal in terms of ISI and ICI effects. Moreover, the majority of the existing studies (e.g., \cite{Gaudio_MIMO_OTFS_Hybrid,MIMO_OTFS_Radar_2020,otfs_radar_2019,OTFS_RadCom_TWC_2020,beamspaceMIMO_OTFS_2022,ISAC_OTFS_JSAC_2022,OTFS_ISAC_part3_2022}) suffers from standard ambiguity limits in range and velocity estimation (i.e., dictated by subcarrier spacing) as far-away and high-mobility targets lead to severe ISI and ICI effects. Recently, several approaches have been proposed to surpass the standard limits in radar detection \cite{OTFS_IOT_2021,ofdm_otfs_comparison_2022}. In \cite{OTFS_IOT_2021}, a virtual CP insertion technique has been considered to increase maximum unambiguous range under the assumption of known number of targets. In \cite{ofdm_otfs_comparison_2022}, improvements in the unambiguous range and velocity of an OTFS correlation receiver have been investigated in a single-target scenario. However, the studies \cite{OTFS_IOT_2021,ofdm_otfs_comparison_2022} share two major shortcomings limiting their applicability and potential in practical OTFS systems. First, these approaches do not address the problem of \textit{multi-target detection with OTFS under ISI and ICI effects}, which could be extremely challenging due to increased side-lobe levels and the accompanying masking effect \cite{MIMO_OFDM_ICI_JSTSP_2021,OTFS_ICC_Workshop_2021}. Second, possible benefits of ISI and ICI in OTFS sensing (e.g., improved target resolvability in range and velocity \cite{MIMO_OFDM_ICI_JSTSP_2021}) have not been explored in \cite{OTFS_IOT_2021,ofdm_otfs_comparison_2022}.

An important aspect of OTFS ISAC schemes pertains to the utilization of multiple-input multiple-output (MIMO) architectures to provide additional degrees-of-freedom in transmission and generate angle estimates together with delay-Doppler measurements in radar sensing \cite{Gaudio_MIMO_OTFS_Hybrid,beamspaceMIMO_OTFS_2022,ISAC_OTFS_JSAC_2022,MIMO_OTFS_Radar_2020,isac_otfs_jstsp_2021}. In \cite{Gaudio_MIMO_OTFS_Hybrid,MIMO_OTFS_Radar_2020,beamspaceMIMO_OTFS_2022}, considering a hybrid analog-digital architecture for MIMO-OTFS ISAC systems, successive interference cancellation based multi-target detection/estimation algorithms have been proposed. The work in \cite{ISAC_OTFS_JSAC_2022} develops a MIMO-OTFS ISAC transmission strategy based on spatial spreading and de-spreading, and investigates the impact of antenna power allocation on communication and sensing performances. Moreover, \cite{isac_otfs_jstsp_2021} designs a sensing-assisted predictive beamforming scheme for MIMO-OTFS ISAC systems in vehicular networks. Despite a significant body of research concerning single-antenna and MIMO OTFS ISAC systems \cite{Gaudio_MIMO_OTFS_Hybrid,otfs_radar_2019,OTFS_RadCom_TWC_2020,MIMO_OTFS_Radar_2020,OTFS_IOT_2021,beamspaceMIMO_OTFS_2022,ISAC_OTFS_JSAC_2022,OTFS_ISAC_part3_2022,ofdm_otfs_comparison_2022,isac_otfs_jstsp_2021}, the following fundamental topics remain unexplored so far: \textit{(i)} development of insightful OTFS signal models for radar and communication that explicitly capture ISI and ICI effects, \textit{(ii)} design of multi-target detection and delay-Doppler-angle estimation algorithms for MIMO-OTFS radar sensing in the presence of ISI and ICI effects, \textit{(iii)} exploitation of ISI and ICI to improve sensing performance of MIMO-OTFS, and \textit{(iv)} investigation and optimization of MIMO-OTFS ISAC trade-offs in spatial and DD domains under varying channel characteristics.



\begin{figure*}
	\centering
	\includegraphics[width=1\linewidth]{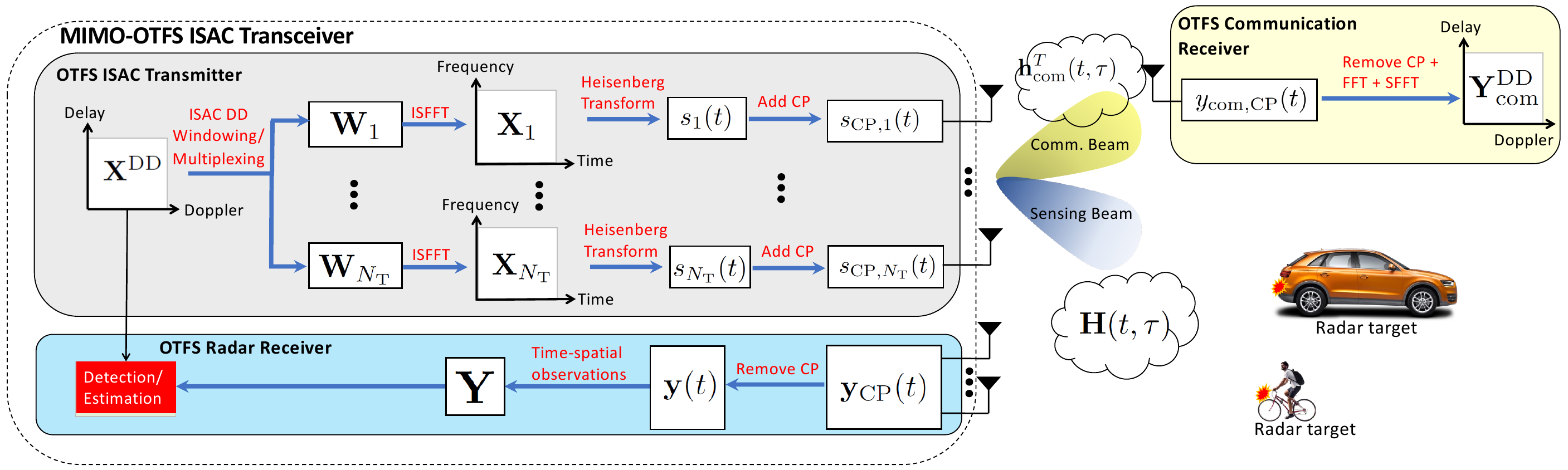}
    \vspace{-6mm}
	\caption{The proposed MIMO-OTFS ISAC system consisting of a multiple-antenna OTFS transmitter and a multiple-antenna OTFS radar receiver on the same hardware platform, and a remote single-antenna OTFS communication receiver. The delay-Doppler data $\boldXdd \in \complexset{N}{M}$ is first windowed using different windowing matrices for each transmit antenna, then converted to the time-frequency domain, and then to the time-domain, where pulse shaping occurs and a single CP is added to the entire frame. The radar receiver detects the objects in the environment using the backscattered signals $\boldY$ given the knowledge of $\boldXdd$, while the communications receiver aims to decode $\boldXdd$ from the received delay-Doppler domain symbols $\boldYcomdd$ given an estimate of $\hhcom(t,\tau)$.}
	\label{fig_system}
\vspace{-0.2in}
\end{figure*}


\subsection{Contributions}
Extending the preliminary results in \cite{OTFS_ICC_Workshop_2021}, this study aims to fill the aforementioned knowledge gaps and proposes a MIMO-OTFS ISAC system by introducing novel observation models, transmission strategies, signal designs and radar receiver algorithms. 
The main contributions can be summarized as follows:
\begin{itemize}
    \item \textbf{Novel Signal Models for OTFS ISAC:} We derive novel radar and communication signal models for MIMO-OTFS ISAC systems by rigorously taking into account ISI and ICI effects. We formulate radar and communication channels as a function of continuous-valued physical path parameters (i.e., delays, Dopplers and angles) to reveal the explicit impact of ISI and ICI on the final observations, providing valuable analytical insights into the manifestation of these effects in OTFS systems.

    \item \textbf{MIMO-OTFS Radar Sensing under ISI/ICI:} Based on the new radar model, we design a generalized likelihood ratio test (GLRT) for multi-target detection/estimation algorithm at the MIMO-OTFS radar receiver that enables simultaneous \textit{mitigation} and \textit{exploitation} of ISI and ICI effects. This approach surpasses the range/velocity ambiguity barrier encountered in most existing OTFS studies \cite{Gaudio_MIMO_OTFS_Hybrid,MIMO_OTFS_Radar_2020,otfs_radar_2019,OTFS_RadCom_TWC_2020,beamspaceMIMO_OTFS_2022,ISAC_OTFS_JSAC_2022,OTFS_ISAC_part3_2022} and allows for detection of any practically relevant range/velocity.

    \item \textbf{MIMO-OTFS ISAC Transmission Schemes:} We propose an adaptive MIMO-OTFS ISAC transmission strategy that considers the different operational modes (i.e., search and track \cite{jointRadCom_review_TCOM}) of the proposed ISAC system. In search mode (where no information is available on sensing and communication directions), we introduce the concept of \textit{delay-Doppler (DD) multiplexing}, which assigns non-overlapping DD bins to TX antennas, enabling omnidirectional transmission from the ISAC transmitter and construction of a virtual array with improved angular resolution at the radar receiver.

    \item \textbf{MIMO-OTFS ISAC Trade-off Optimization:} For the track mode, we develop an algorithm for  ISAC signal design that optimizes the trade-off between radar and communications in the presence of a-priori location information on radar targets and the communication receiver. We derive an achievable rate expression based on the covariance matrix of the linear minimum mean square error (LMMSE) estimator and formulate an ISAC trade-off problem by optimizing over DD domain power allocation and transmit beamforming, following a directional phased-array transmission strategy.

\end{itemize}


\textit{Notations:} $\diag{\xx}$ outputs a diagonal matrix with the elements of a vector $\xx$ on the diagonals, $\diag{\boldX}$ represents a diagonal matrix with the diagonal elements of a square matrix $\boldX$ on the diagonals, $\vecc{\cdot}$ denotes matrix vectorization operator, and $\veccinv{\cdot}$ reshapes a vector into an $N \times M$ matrix column-wise. $\odot$ and $\otimes$ denote the Hadamard (element-wise) and Kronecker products, respectively. $\norm{\boldX}_F$ is the Frobenius norm of $\boldX$.

\section{MIMO-OTFS ISAC System Model}
We consider a MIMO-OTFS ISAC system consisting of a MIMO-OTFS ISAC transceiver and a single-antenna OTFS communication receiver (RX), as shown in Fig.~\ref{fig_system}. The transceiver contains, on the same hardware platform, \textit{(i)} an ISAC transmitter (TX) with an $\Ntx$-element digital array that generates the OTFS ISAC signal for sending data symbols to the communication RX, and \textit{(ii)} a radar RX with an $\Nrx$-element digital array that processes the backscattered signals for target detection and parameter estimation \cite{OTFS_RadCom_TWC_2020}. In this section, we present the proposed MIMO-OTFS transmission scheme, and derive the OTFS transmit signal model and the corresponding observation models at the radar and communication receivers by adopting the single cyclic-prefix (CP) OTFS modulation architecture \cite{reducedCP_OTFS_2018,otfs_radar_2019,otfs_frac_2020,OTFS_RadCom_TWC_2020,OTFS_Eq_Learning_TWC_2022,ISAC_OTFS_JSAC_2022}. 



\begin{figure}
	\centering
	\includegraphics[width=0.4\linewidth]{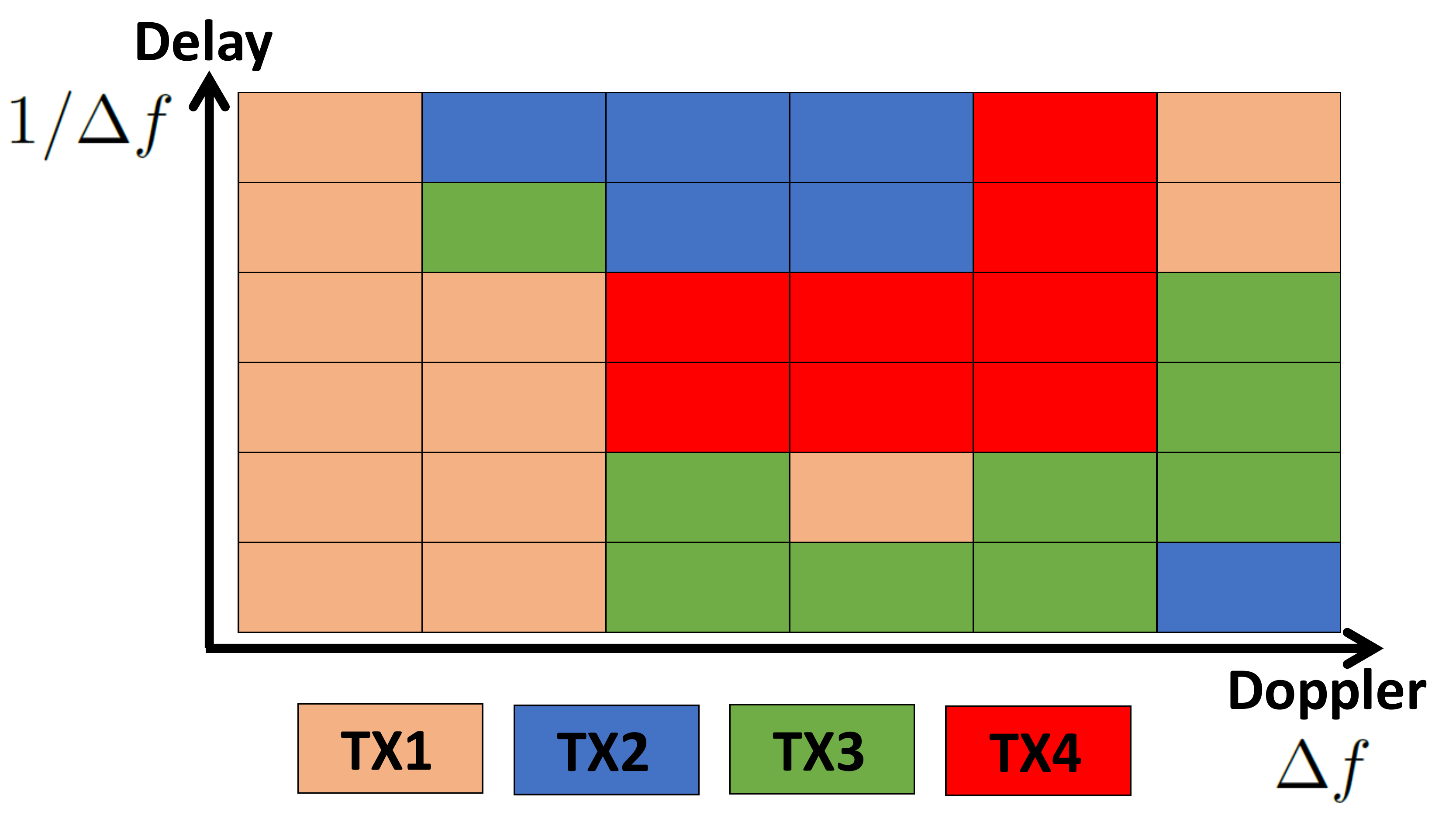}
 \vspace{-0.1in}
	\caption{MIMO-OTFS system with delay-Doppler (DD) domain multiplexing of symbols to transmit antennas.}
	\label{fig_dd_multiplex}
	\vspace{-0.2in}
\end{figure}

\subsection{OTFS Transmit Signal Model}\label{sec_otfs_tx}
The OTFS system has a total bandwidth $N \deltaf$ and total frame duration $M T$ (excluding any CP), where $N$ and $M$ denote the number of subcarriers and the number of symbols, respectively, $\deltaf$ is the subcarrier spacing and $T = 1/\deltaf$ represents the symbol duration. Let $\boldXdd \in \complexset{N}{M}$ denote the two-dimensional (2-D) OTFS frame in the DD domain consisting of $NM$ transmit data symbols that reside on the DD grid 
\begin{equation}\nonumber
    \mathcal{G} = \left\{ \left(\frac{n}{N \deltaf}, \frac{m}{MT}\right) ~\Big\lvert ~ 0 \leq n \leq N-1, 0 \leq m \leq M-1 \right\}~.
\end{equation}
In the following, $\boldXdd$ will be processed in four steps: \textit{(i)} DD windowing/precoding of $\boldXdd$ for transmission over multiple TX antennas, \textit{(ii)} transforming from DD to frequency-time (FT) domain, \textit{(iii)} Heisenberg transform, and \textit{(iv)} addition of CP.

\subsubsection{Delay-Doppler Windowing/Precoding}\label{sec_dd_window}
In the proposed DD windowing scheme, the $\thn{i}$ TX antenna transmits a windowed/precoded version of the OTFS frame $\boldXdd$, represented by \cite{ISAC_OTFS_JSAC_2022}
\begin{align} \label{eq_xi_dd}
    \boldXdd_i &\triangleq \boldXdd \odot \boldW_i~,
\end{align}
where $\boldW_i \in \complexset{N}{M}$ is the windowing matrix for the $\thn{i}$ antenna satisfying the total power constraint $\sum_{i=1}^{\Ntx} \norm{\boldW_i}_F^2 = NM \Ntx$.
Note that the DD windowing operation in \eqref{eq_xi_dd} can be interpreted as \textit{symbol-wise precoding} in the DD domain \cite{ISAC_OTFS_JSAC_2022}. Depending on the operational mode of the OTFS ISAC system, the windowing matrices $\{\boldW_i\}_{i=1}^{\Ntx}$ can be designed in different ways. 
\begin{itemize}
    \item \textit{Search Mode:} In search mode where no \textit{a-priori} information about radar targets and the communication RX is available, the ISAC TX can follow an omnidirectional transmission strategy \cite{jointRadCom_review_TCOM} and achieve orthogonality across the TX antennas by dividing the entire DD domain into mutually exclusive and collectively exhaustive subsets of $\Ntx$ Boolean masks (see Fig.~\ref{fig_dd_multiplex}), i.e., 
    \begin{align} \label{eq_dd_mult}
    \boldW_i \odot \boldW_j = \boldzero_{N \times M}\,, \, \forall i \neq j \,, ~~    
    \sum_{i=1}^{\Ntx} \boldW_i = \boldone_{N \times M} ~.
    \end{align}
    The proposed search mode strategy in \eqref{eq_dd_mult} is hereafter called \textit{DD multiplexing}. This enables constructing a virtual array of $\Ntx \Nrx$ elements for MIMO-OTFS radar sensing to improve angular resolvability of multiple targets, as will be elaborated on in Sec.~\ref{sec_search_DD_mult}. 
    
    \item \textit{Track Mode:} Once the communication RX and/or targets are detected in search mode\footnote{Search and track modes can be interpreted in the context of sensing-assisted beam management (or, \textit{sensing-assisted communications} \cite{sensingAssistedComm_TWC_2020,ISAC_V2I_Extended_TWC_2022}). First, the communication RX can be sensed in search mode as a passive object using the echoes of the downlink transmission and then distinguished from other targets in the environment by the ISAC TX in the uplink phase \cite{jointRadCom_review_TCOM}.}, the ISAC TX acquires \textit{a-priori} knowledge on the communication and sensing directions. Then, the ISAC TX can switch to track mode and construct the windows to steer beams towards desired angular sectors \cite{multibeam_TVT_2019,DFRC_Waveform_Design,JCR_JSTSP_2021,beamformer_ISAC_FD_TCOM_2022}. In track mode, we transmit the same waveform over all antennas (i.e., phased-array radar \cite{phasedMIMO_radar_TSP_2010}) with adjustable complex coefficients per antenna to maximize SNR towards an angular sector of interest. This also enables optimizing the trade-off between radar and communications. In compliance with the phased-array transmission, we can express the windows in \eqref{eq_xi_dd} as
\begin{align} \label{eq_wwi_phased}
    \ww_i = \beta_i \pp ~,
\end{align}
where $\ww_i \triangleq \vecc{\boldW_i} \in \complexset{NM}{1}$, $\pp \in \realset{NM}{1}$ contains the amplitudes for the DD bins, common to all TX antennas, and $\beta_i \in \complexsett$ denotes the complex coefficient applied at the $\thn{i}$ TX antenna, used for beam steering. The design of $\{\beta_i\}_{i=1}^{\Ntx}$ and $\pp$ will be discussed in Sec.~\ref{sec_des_track}. 
\end{itemize}


\subsubsection{Transformation from Delay-Doppler to Frequency-Time Domain}
Applying an inverse symplectic finite Fourier transform (ISFFT) (i.e., an $N$-point FFT over the columns and an $M$-point IFFT over the rows of $\boldXdd$), we transform the 2-D transmit data block from the DD domain to the frequency-time (FT) domain \cite{hadani2017orthogonal,OTFS_CE_TSP_2019,reducedCP_OTFS_2018,OFDM_OTFS_modem_2017,OTFS_Canc_Iterative_TWC_2018}
\begin{equation}\label{eq_otfs_dd2ft}
    \boldX_i = \FF_N \boldXdd_i \FF_M^H ~, 
\end{equation}
where $\boldX_i \in \complexset{N}{M}$ is the FT domain signal for the $\thn{i}$ TX antenna and $\FF_N \in \complexset{N}{N}$ is a unitary discrete Fourier transform (DFT) matrix with $\left[ \FF_N\right]_{\ell,n} = \frac{1}{\sqrt{N}} e^{- j 2 \pi n \frac{\ell}{N}} $. Note that the DD domain windowing in \eqref{eq_xi_dd} can equivalently be implemented as a 2D filter in the FT domain since point-wise multiplication in the DD domain corresponds to 2D circular convolution in the FT domain \cite{windowDesign_OTFS_TCOM_2021}.

\subsubsection{Heisenberg Transform}
To map the FT domain 2-D sequence $\boldX_i$ to a time domain signal transmitted over the wireless channel, we apply the \textit{Heisenberg transform} \cite{hadani2017orthogonal,OTFS_Canc_Iterative_TWC_2018}, which entails an $N$-point IFFT together with a transmit pulse-shaping waveform $\gtx(t)$ (which is time limited to $\left[0, \, T \right]$). The time domain signal for the $\thn{i}$ TX antenna and the $\thn{m}$ symbol after the Heisenberg transform can be written as
\begin{equation}\label{eq_smt}
    s_{m,i}(t) = \frac{1}{\sqrt{N}} \sum_{n = 0}^{N-1}  [\boldX_i]_{n,m} \, e^{j 2 \pi n \deltaf t} \gtx(t) \,, ~ 0 \leq t \leq T ~.
\end{equation}
Hence, the time domain signal transmitted by the $\thn{i}$ TX antenna for the entire OTFS frame without CP is given by
\begin{equation}\label{eq_st}
    s_i(t) = \sum_{m=0}^{M-1} s_{m,i}(t-mT) \, , ~ 0 \leq t \leq MT ~.
\end{equation}
\subsubsection{CP Addition}
Finally, the entire time domain signal with CP  (of duration $\Tcp$)  for the $\thn{i}$ TX antenna is given by \cite{reducedCP_OTFS_2018,otfs_radar_2019,otfs_frac_2020,OTFS_RadCom_TWC_2020,OTFS_Eq_Learning_TWC_2022}
\begin{equation}\label{eq_smaone_baseband}
\scpi(t) = \begin{cases} s_i(t) ,&~~ 0 \leq t \leq MT \\
s_i(t+MT) ,&~~ -\Tcp \leq t \leq 0 \end{cases}  ~.
\end{equation} 


\subsection{OTFS Radar Signal Model}\label{sec_rec_model}
Assuming the existence of $K$ targets in the sensing environment, we consider the narrowband time-varying MIMO radar channel model \cite{80211_Radar_TVT_2018,OTFS_RadCom_TWC_2020,MIMO_OTFS_Radar_2020,beamspaceMIMO_OTFS_2022} 
\begin{align}\label{eq_mimo_channel}
    \boldH(t, \tau) = \sum_{k=0}^{K-1} \alpha_k \delta(\tau - \tau_k)  e^{j 2 \pi \nu_k t} \arx(\theta_k) \atx^T(\theta_k) ~,
\end{align}
where $\atx(\theta) \in \complexset{\Ntx}{1}$ and $\arx(\theta) \in \complexset{\Nrx}{1}$ denote the steering vectors of the TX and radar RX arrays, respectively, and the $\thn{k}$ target is characterized by a complex channel gain $\alpha_k$, a round-trip delay $\tauk = 2 R_k/c$, a Doppler shift $\nuk = 2  \vk/ \lambda$ and an \ac{AOA}/\ac{AOD}\footnote{Due to co-located TX and radar RX arrays, \ac{AOA} and \ac{AOD} assume the same value for the radar channel.} $\theta_k$, with $R_k$, $\vk$, $c$ and $\lambda$ denoting the range, radial velocity, speed of propagation and carrier wavelength, respectively. Based on the channel model in \eqref{eq_mimo_channel}, the backscattered signal at the RX array of the  radar receiver is 
\begin{align} 
    \yycp(t) \label{eq_yt} 
    &= \int \boldH(t, \tau) \ssscp(t-\tau) \d \tau +\zz(t)  = \sum_{k=0}^{K-1} \alpha_k e^{j 2 \pi \nu_k t} \arx(\theta_k) \atx^T(\theta_k) \ssscp(t - \tau_k)  +\zz(t) \in \complexset{\Nrx}{1}
\end{align}
for $-\Tcp \leq t \leq M T$, where $\ssscp(t) = [ \scpig{1}(t) \ \cdots \ \scpig{\Ntx}(t) ]^T \in \complexset{\Ntx}{1}$, 
with $\scpi(t)$ in \eqref{eq_smaone_baseband}, and $\zz(t) \in \complexset{\Nrx}{1} $ is additive white Gaussian noise (AWGN).

\subsection{OTFS Communication Signal Model}\label{sec_comm_model}
Suppose that the communication channel between the ISAC TX and the communication RX consists of $\Ktilde$ paths. Then, similar to the radar channel in \eqref{eq_mimo_channel}, the \ac{MISO} OTFS communication channel can be modeled as \cite{OTFS_RadCom_TWC_2020,ISAC_OTFS_JSAC_2022}
\begin{align}\label{eq_miso_channel_comm}
    \hhcom^T(t, \tau) = \sum_{k=0}^{\Ktilde-1} \alphat_k \delta(\tau - \taut_k)  e^{j 2 \pi \nut_k t}  \atx^T(\thetat_k) ~,
\end{align}
where $\alphat_k$, $\taut_k$, $\nut_k$ and $\theta_k$ represent, respectively, the complex channel gain, delay, Doppler shift and \ac{AOD} of the $\thn{k}$ path. Using \eqref{eq_smaone_baseband} and \eqref{eq_miso_channel_comm}, the received signal at the OTFS communication RX is given by
\begin{align} 
    \ycomcp(t) \label{eq_yt_com} 
    &= \int \hhcom^T(t, \tau) \ssscp(t-\tau) \d \tau + \ztilde(t)   = \sum_{k=0}^{\Ktilde-1} \alphat_k e^{j 2 \pi \nut_k t} \atx^T(\thetat_k) \ssscp(t - \taut_k)  + \ztilde(t)
\end{align}
for $-\Tcp \leq t \leq M T$, where $\ztilde(t)$ denotes AWGN. To prevent inter-frame interference for both radar and communications, the CP duration is assumed to be larger than the round-trip delay of the furthermost target \cite{Firat_OFDM_2012,OFDM_Radar_Phd_2014,SPM_JRC_2019,MIMO_OFDM_ICI_JSTSP_2021} and the delay spread of the communication channel \cite{raviteja2018practical,OTFS_CE_TSP_2019}.

\section{Novel OTFS Signal Model Accounting for ISI/ICI Effects}\label{sec_ma1_model}
In this section, we derive a novel compact representation of the OTFS radar and communication signal models in \eqref{eq_yt} and \eqref{eq_yt_com}, respectively, which rigorously captures the ISI and ICI effects \cite{OTFS_Canc_Iterative_TWC_2018,OTFS_CE_TSP_2019,OTFS_RadCom_TWC_2020,beamspaceMIMO_OTFS_2022}. We demonstrate that the new formulation provides important insights into the manifestation of the ISI and ICI effects, and enables their exploitation to improve OTFS radar performance.

\subsection{Time-Spatial Observations for OTFS Radar}\label{sec_der_rec_otfs}
\subsubsection{CP Removal}
We begin by formulating the OTFS radar RX signal in \eqref{eq_yt} in a compact form. First, we remove the CP in \eqref{eq_yt} (i.e., the interval $\left[-\Tcp, \, 0 \right]$) to obtain
\begin{equation}\label{eq_yt_baseband}
\yy(t) = \begin{cases} \yycp(t) ,&~~ 0 \leq t \leq MT \\
\boldzero ,&~~ -\Tcp \leq t \leq 0 \end{cases}  ~.
\end{equation} 
Under the assumption $\Tcp \geq \max_k \tau_k$, \eqref{eq_smaone_baseband} implies the cyclic shift property \cite[Eq.~(6)]{reducedCP_OTFS_2018}
\begin{align}
    s_i([t - \tau_k]_{MT}) = \scpi(t - \tau_k), ~ 0 \leq t \leq MT ~,
\end{align}
where $[\cdot]_T$ denotes modulo-$T$. Accordingly, using \eqref{eq_yt}, the signal in \eqref{eq_yt_baseband} becomes 
\begin{align}\label{eq_yt_ma1_cp}
        \yy(t) 
        &= \sum_{k=0}^{K-1} \alpha_k e^{j 2 \pi \nu_k t} \arx(\theta_k) \atx^T(\theta_k) \sss([t - \tau_k]_{MT})  +\zz(t) ~,
\end{align}
where 
\begin{align}\label{eq_st_vec}
    \sss(t) = [ s_1(t) \ \cdots \ s_{\Ntx}(t) ]^T \in \complexset{\Ntx}{1} ~,
\end{align}
with $s_i(t)$ being defined in \eqref{eq_st}.

\subsubsection{Frequency-Domain Representation}
Let $S_i(f) \triangleq \mathcal{F}\{s_i(t)\} = \int_{0}^{MT} s_i(t) e^{-j 2 \pi f t}\d t$ denote the Fourier transform of $s_i(t)$. Then, a cyclic shift of $s_i(t)$ corresponds to a phase shift in $S_i(f)$, i.e.,
\begin{equation} \label{eq_si_ft}
    s_i([t - \tau_k]_{MT}) = \mathcal{F}^{-1}\big\{ S_i(f) e^{-j 2 \pi f \tau_k} \big\} ~,
\end{equation}
where $\mathcal{F}^{-1}\{\cdot\}$ represents the inverse Fourier transform. Sampling the time domain at $t = \ell T/N$ for $\ell = 0, \ldots, NM-1$ and the frequency domain at $f = n \deltaf/M$ for $n = 0, \ldots, NM-1$, the equivalent discrete-time representation of \eqref{eq_si_ft} can be written as
\begin{align} \label{eq_si_cyclic}
    s_i([t - \tau_k]_{MT}) \big\lvert_{t=\ell T/N} = \FF^H \big( \FF \sss_i \odot \bb(\tau_k) \big) 
\end{align}
for $\ell = 0, \ldots, NM-1$, where $\FF \in \complexset{NM}{NM}$ is a unitary DFT matrix and $\sss_i \in \complexset{NM}{1}$ denotes the sampled version of $s_i(t)$ in \eqref{eq_st_vec}. Note that $\sss_i$ can be expressed using \eqref{eq_otfs_dd2ft}--\eqref{eq_st} as \cite{raviteja2018practical}
\begin{align} \label{eq_sssi}
    \sss_i = \vecc{\gtxmat \FF_N^H \boldX_i} = \vecc{\gtxmat \boldXdd_i \FF_M^H} ~, 
\end{align}
where $\gtxmat \triangleq \diag{ \gtx(0), \gtx(T/N), \ldots, \gtx((N-1)T/N) }$. Additionally,
\begin{equation}\label{eq_b_steer}
    \bb(\tau) = \bb_{N}(\tau) \otimes \bbisi(\tau) \in \complexset{NM}{1}
\end{equation}
is the \textit{frequency-domain steering vector} with $\bb_{N}(\tau) \triangleq  \transpose{ \left[ 1 \ e^{-j 2 \pi \deltaf \tau} \cdots \  e^{-j 2 \pi (N-1) \deltaf  \tau} \right] } \in \complexset{N}{1}$ and $ \bbisi(\tau) \triangleq  \transpose{ \left[ 1 \ e^{-j 2 \pi \frac{1}{M} \deltaf \tau} \ \cdots \  e^{-j 2 \pi \frac{M-1}{M} \deltaf  \tau} \right] } \in \complexset{M}{1}$. 
To express the Doppler-dependent term in \eqref{eq_yt_ma1_cp} in a compact manner, let us define the \textit{temporal steering vector}
\begin{equation}\label{eq_c_steer}
    \cc(\nu) = \cc_{M}(\nu) \otimes \ccici(\nu) \in \complexset{NM}{1}    
\end{equation}
with $ \cc_M(\nu)  \triangleq \transpose{ \left[ 1 \ e^{j 2 \pi T \nu } \ \ldots \  e^{j 2 \pi (M-1) T \nu } \right] } \in \complexset{M}{1}$ and $ \ccici(\nu) \triangleq \transpose{ \left[ 1 \ e^{j 2 \pi  \frac{T}{N} \nu} \ \ldots \ e^{j 2 \pi \frac{T(N-1)}{N} \nu}  \right]   } \in \complexset{N}{1}$.

\subsubsection{Time-Spatial Observations}
Using \eqref{eq_st_vec}, \eqref{eq_si_cyclic}, \eqref{eq_b_steer} and \eqref{eq_c_steer}, the time-varying terms in \eqref{eq_yt_ma1_cp} can be written for $\ell = 0, \ldots, NM-1$ as
\begin{align} \nonumber
    & e^{j 2 \pi \nu_k t} \atx^T(\theta_k) \sss([t - \tau_k]_{MT}) \big\lvert_{t=\ell T/N}
     = \left( \sum_{i=1}^{\Ntx}   \FF^H \big( \FF \sss_i \odot \bb(\tau_k) \big) \left[ \atx(\theta_k) \right]_i \right) \odot \cc(\nu_k)  ~,
    \\  &=  \boldC(\nu_k) \sum_{i=1}^{\Ntx} \FF^H \boldB(\tau_k) \FF \sss_i \left[ \atx(\theta_k) \right]_i 
    = \boldC(\nu_k)  \FF^H \boldB(\tau_k) \FF \boldS  \atx(\theta_k) \in \complexset{NM}{1}\label{eq_time_compact} ~,
\end{align}
where $\boldB(\tau) \triangleq \diag{\bb(\tau)} \in \complexset{NM}{NM}$, $\boldC(\nu) \triangleq \diag{\cc(\nu)} \in \complexset{NM}{NM}$, and
\begin{align} \label{eq_s_tx}
    \boldS \triangleq \left[ \sss_1 \, \ldots \, \sss_{\Ntx} \right] \in \complexset{NM}{\Ntx}
\end{align}
is the transmit waveform matrix, i.e., the sampled version of \eqref{eq_st_vec}. Plugging \eqref{eq_time_compact} into \eqref{eq_yt_ma1_cp}, the sampled observations arranged into time-spatial form are given, for $\ell = 0, \ldots, NM-1$, by
\begin{align}\label{eq_obs_mimo}
    \boldY \triangleq\yy(t) \big\lvert_{t = \ell T/N}   = \sum_{k=0}^{K-1} \alpha_k     \boldC(\nu_k)  \FF^H \boldB(\tau_k) \FF \boldS  \atx(\theta_k) \arx^T(\theta_k) + \boldZ \in \complexset{NM}{\Nrx} ~,
\end{align}
where $\boldZ \in \complexset{NM}{\Nrx}$ is the additive noise matrix with $\vecc{\boldZ} \sim \mtCN(\boldzero, \sigma^2 \Imatrix )$. We note that the radar model \eqref{eq_obs_mimo} is valid for any single-CP waveform of the form \eqref{eq_smaone_baseband}.


\subsection{Delay-Doppler Observations for OTFS Communications}
Since the communication model in \eqref{eq_yt_com} has the same structure as the radar model in \eqref{eq_yt}, we can follow the same steps as applied for the radar side in the previous part. Accordingly, the time-domain observations at the communication RX after CP removal can be obtained from \eqref{eq_yt_com} for $\ell = 0, \ldots, NM-1$ as
\begin{align}\label{eq_obs_miso_com}
    \yycom \triangleq \ycomcp(t) \big\lvert_{t = \ell T/N}   = \sum_{k=0}^{\Ktilde-1} \alphat_k     \boldC(\nut_k)  \FF^H \boldB(\taut_k) \FF \boldS  \atx(\thetat_k)  + \zztilde \in \complexset{NM}{1} ~,
\end{align}
where $\zztilde \in \complexset{NM}{1}$ is the AWGN vector with $\zztilde \sim \mtCN(\boldzero, \sigma^2 \Imatrix )$. The DD domain symbols can be obtained from the time-domain observations in \eqref{eq_obs_miso_com} by inverting the transmit side operations in Sec.~\ref{sec_otfs_tx} as \cite{raviteja2018practical,windowDesign_OTFS_TCOM_2021,ISAC_OTFS_JSAC_2022}
\begin{align} \label{eq_ycomdd}
    \boldYcomdd = \FF_N^H \big( \FF_N \grxmat \boldYcom \big) \FF_M \in \complexset{N}{M} ~,
\end{align}
where $\boldYcom \triangleq \veccinv{\yycom} \in \complexset{N}{M}$ and $\grxmat = \diag{ \grx(0), \grx(T/N), \ldots, \grx((N-1)T/N) }$ with $\grx(t)$ denoting the pulse shaping filter at the receiver. In \eqref{eq_ycomdd}, we apply receive pulse shaping for each symbol (i.e., over the columns of $\boldYcom$), take an $N$-point FFT over the columns to switch to the FT domain and take an SFFT (i.e., an $N$-point IFFT over the columns and an $M$-point FFT over the rows) to switch from the FT domain to the DD domain \cite{OTFS_CE_TSP_2019}. 

Assuming rectangular pulse shaping $\gtxmat = \grxmat = \Imatrix_N$ \cite{raviteja2018practical,ISAC_OTFS_JSAC_2022}, the vectorized version of \eqref{eq_ycomdd} can be written as \cite{raviteja2018practical}
\begin{align}\label{eq_ycomdd_vec}
    \yycomdd = (\FF_M \otimes \Imatrix_N) \yycom \in \complexset{NM}{1} ~,
\end{align}
where $\yycomdd \triangleq \vecc{\boldYcomdd}$. Inserting \eqref{eq_obs_miso_com} into \eqref{eq_ycomdd_vec} yields\footnote{We keep the same notation for the noise vector for ease of exposition as the noise statistics do not change via multiplication by the unitary matrix $\FF_M \otimes \Imatrix_N $.}
\begin{align} \label{eq_yycom_dd2}
    \yycomdd &= (\FF_M \otimes \Imatrix_N) \sum_{k=0}^{\Ktilde-1} \alphat_k     \boldC(\nut_k)  \FF^H \boldB(\taut_k) \FF \boldS  \atx(\thetat_k) + \zztilde ~.
\end{align}
To make explicit the relation between the transmit DD domain symbols $\boldXdd$ in \eqref{eq_xi_dd} and the received DD domain symbols in \eqref{eq_yycom_dd2}, we re-write $\boldS  \atx(\thetat_k)$ as
\begin{align} \label{eq_sat_1}
    \boldS  \atx(\thetat_k) &= \sum_{i=1}^{\Ntx} \sss_i [ \atx(\thetat_k)]_i = \sum_{i=1}^{\Ntx} (\FF_M^H \otimes \Imatrix_N) \xxdd_i [\atx(\thetat_k)]_i ~,
    \\ \label{eq_sat_2}
    &= (\FF_M^H \otimes \Imatrix_N) \sum_{i=1}^{\Ntx}  (\xxdd \odot \ww_i) [\atx(\thetat_k)]_i 
    = (\FF_M^H \otimes \Imatrix_N) \big( \xxdd \odot \boldW \atx(\thetat_k) \big) ~,
    \\ \label{eq_sat_der}
    &= (\FF_M^H \otimes \Imatrix_N) \diagb{ \boldW \atx(\thetat_k) } \xxdd ~,
\end{align}
where $\xxdd_i \triangleq \vecc{\boldXdd_i} \in \complexset{NM}{1}$, $\xxdd \triangleq \vecc{\boldXdd} \in \complexset{NM}{1}$ and $\boldW \triangleq [\ww_1 \ \cdots \ \ww_{\Ntx}] \in \complexset{NM}{\Ntx}$. Here, \eqref{eq_sat_1} and \eqref{eq_sat_2} follow from \eqref{eq_sssi} and \eqref{eq_xi_dd}, respectively. Then, using \eqref{eq_sat_der}, the received DD symbols in \eqref{eq_yycom_dd2} can be compactly expressed as
\begin{align} \label{eq_yycom_dd_comp}
    \yycomdd &= \boldHdd \xxdd + \zztilde ~,
\end{align}
where the DD domain channel matrix is given by
\begin{align} \label{eq_hdd}
    \boldHdd &= (\FF_M \otimes \Imatrix_N) \sum_{k=0}^{\Ktilde-1} \Big[ \alphat_k     \boldC(\nut_k)  \FF^H \boldB(\taut_k) \FF
     (\FF_M^H \otimes \Imatrix_N) \diagb{ \boldW \atx(\thetat_k) } \Big] \in \complexset{NM}{NM} ~. 
\end{align}

\subsection{Manifestation of ISI and ICI Effects}\label{sec_manifest}
We now elaborate on how ISI and ICI effects manifest themselves in the radar signal model \eqref{eq_obs_mimo} by analyzing the structure of the steering vectors in \eqref{eq_b_steer} and \eqref{eq_c_steer}. The vectors $\bb_{N}(\tau)$ and $\cc_M(\nu)$ commonly arise in the context of OFDM radar, used for recovering target range and velocity, respectively \cite{Passive_OFDM_2010,OFDM_Passive_Res_2017_TSP,OFDM_Radar_Phd_2014,RadCom_Proc_IEEE_2011,OFDM_DFRC_TSP_2021,MIMO_OFDM_ICI_JSTSP_2021}. In contrast, the vectors $\bbisi(\tau)$ and $\ccici(\nu)$ specify disturbances degrading radar performance when standard FFT-based algorithms are employed for range-velocity estimation, e.g., \cite{RadCom_Proc_IEEE_2011,ofdm_radar_correlation_TAES_2020}. 
In particular, 
\begin{itemize}
    \item $\bb_N(\tau)$ quantifies delay-dependent frequency-domain phase rotations corresponding to the Fourier transform of fast-time (hereafter called \textit{fast-frequency} domain), while $\bbisi(\tau)$ involves inter-symbol (slow-time) delay-dependent phase rotations (hereafter called \textit{slow-frequency} domain), leading to \textit{inter-symbol interference (ISI)}.
    \item $\ccici(\nu)$ represents Doppler-induced fast-time phase rotations causing \textit{inter-carrier interference (ICI)}, similar to the carrier frequency offset (CFO) effect in OFDM communications \cite{Visa_CFO_TSP_2006}, while $\cc_M(\nu)$ captures Doppler-dependent slow-time phase progressions.
\end{itemize}
Similar analysis holds true also for the communication model \eqref{eq_yycom_dd_comp}, where the ISI and ICI among the DD symbols $\xxdd$ \cite{OTFS_RadCom_TWC_2020,OTFS_Canc_Iterative_TWC_2018} are captured through $\boldB(\tau)$ and $\boldC(\nu)$ in the DD domain channel \eqref{eq_hdd}.

\subsection{Increasing Unambiguous Detection Intervals for OTFS Radar via ISI and ICI Exploitation}\label{sec_isi_ici}
\begin{figure}[t]
        \begin{center}
        \subfigure[]{
			 \label{fig_b_ISI}
			 \includegraphics[width=0.42\textwidth]{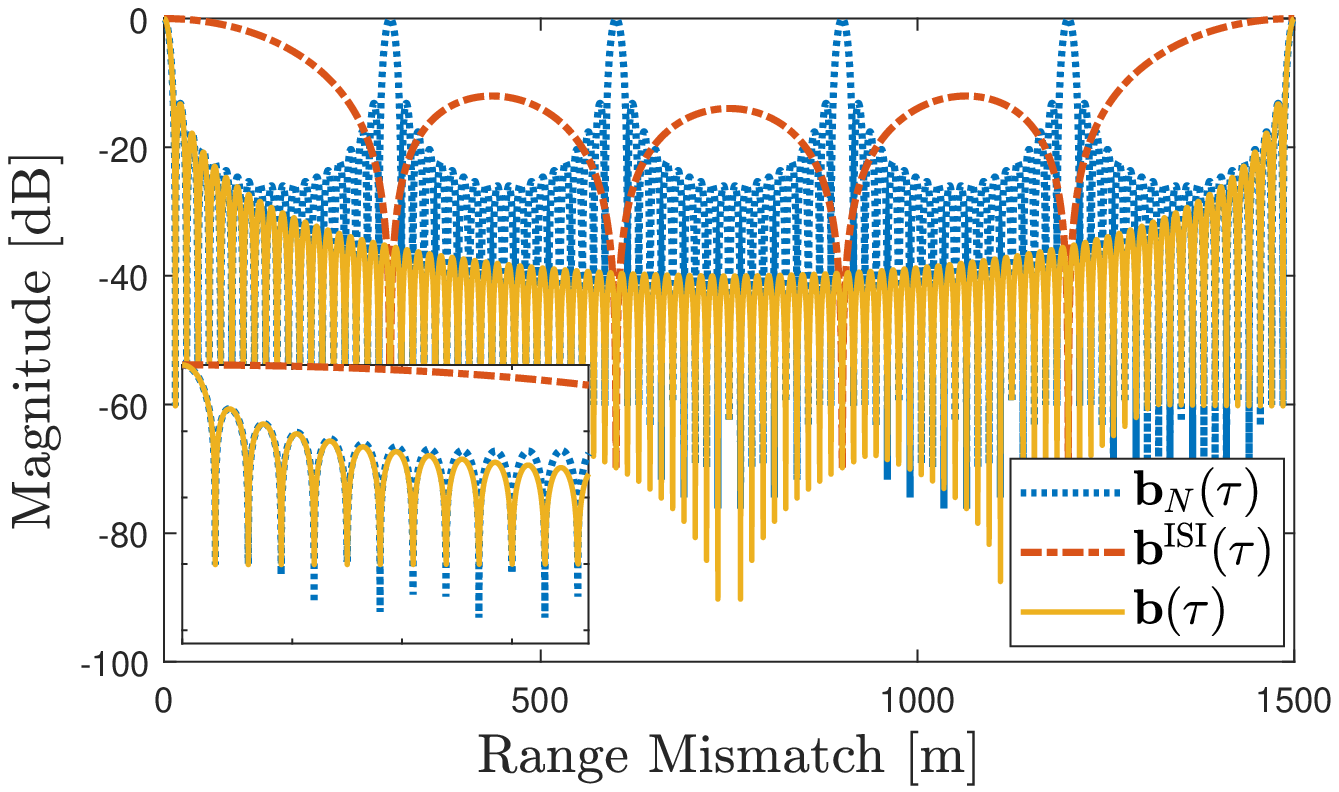}
		}
        \subfigure[]{
			 \label{fig_b_ISI_virtual}
			 \includegraphics[width=0.42\textwidth]{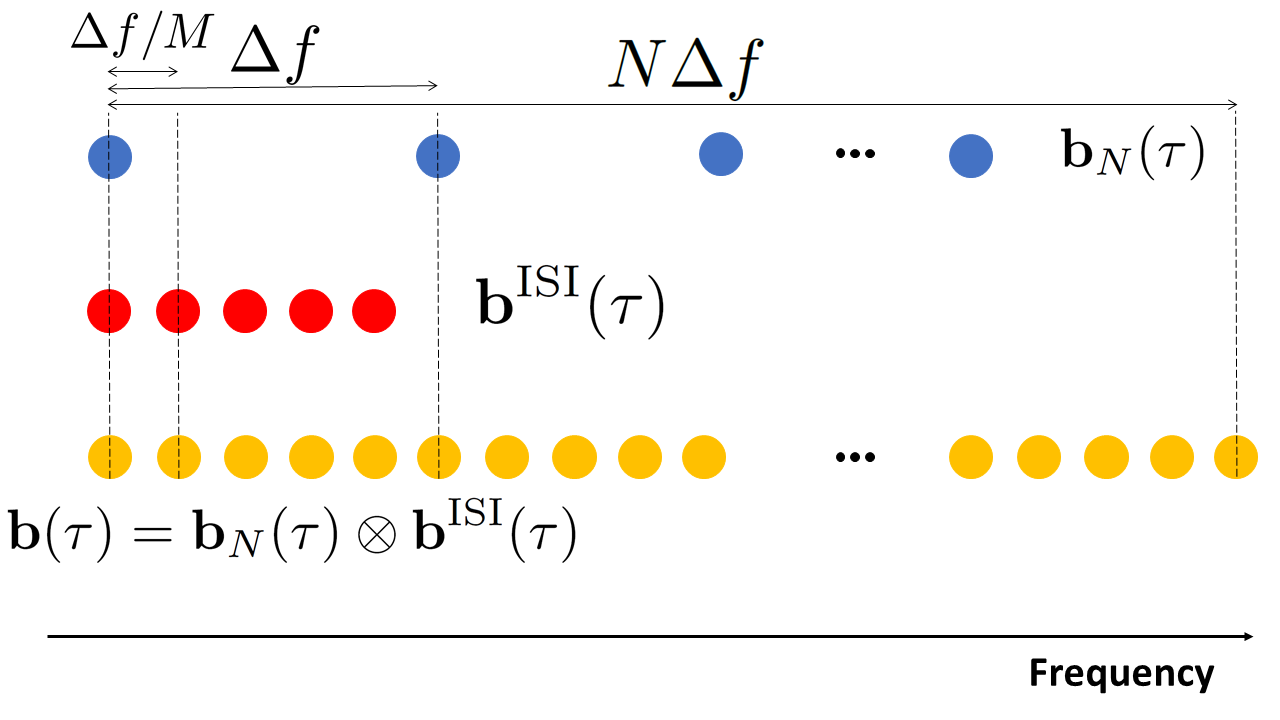}
		}
		\end{center}
		\vspace{-0.2in}
        \caption{\subref{fig_b_ISI} Range profiles corresponding to the OTFS frequency-domain steering vectors in \eqref{eq_b_steer}, obtained by taking the inverse DFT at $\tau = 0$, where $N = 20$, $M=5$ and $\deltaf = 500 \, \rm{kHz}$. The inset shows the zoomed-in version of the first several resolution cells. Similar to MIMO radar with virtual array aperture, the Kronecker structure in \eqref{eq_b_steer} introduced by the ISI effect enables $\bb(\tau)$ to increase the unambiguous range of $\bb_N(\tau)$ by a factor of $M$ while retaining its range resolution, creating a virtual frequency-domain array, as shown in \subref{fig_b_ISI_virtual}.} 
        \label{fig_b_ISI_all}
        \vspace{-0.1in}
\end{figure}

The steering vector structures in \eqref{eq_b_steer} and \eqref{eq_c_steer} suggest that both radar and communication receivers in  OTFS suffer from ISI and ICI effects \cite{OTFS_Canc_Iterative_TWC_2018,OTFS_RadCom_TWC_2020}. However, unlike OTFS communications, these two effects can be turned into an advantage for OTFS radar. In particular, ISI manifests itself through the \textit{slow-frequency} steering vector $\bbisi(\tau)$ and enables sampling the available bandwidth $N \deltaf$ at integer multiples of $\deltaf / M$ as observed from the Kronecker structure in \eqref{eq_b_steer}. In ISI-free operation, the steering vector in \eqref{eq_b_steer} would only involve the \textit{fast-frequency} component $\bb_{N}(\tau)$, in which case the bandwidth can only be sampled with a spacing of $\deltaf$, i.e., the subcarrier spacing. Hence, ISI can increase the maximum detectable unambiguous delay by a factor of $M$ by allowing the frequency domain to be sampled $M$ times faster compared to a standard ISI-free radar operation (e.g., in OFDM-based OTFS radar\footnote{Contrary to single-CP OTFS \cite{ISAC_OTFS_JSAC_2022}, OFDM-based OTFS systems use separate CPs for each symbol in the OTFS/OFDM frame to circumvent the ISI effect \cite{OTFS_CE_TSP_2019,otfs_modem_2018}.}). More precisely, the unambiguous delays with and without ISI are given, respectively, by
\begin{align}\label{eq_isi_exp}
    \taumaxisi = \min \Big\{ \frac{M}{\deltaf}, \Tcp \Big\}, ~ \taumax = \min \Big\{ \frac{1}{\deltaf}, \Tcp \Big\} ~,
\end{align}
where $\Tcp$ is the upper limit to prevent inter-frame interference (i.e., between consecutive OTFS frames) in single-CP OTFS considered in our model \cite{OTFS_IOT_2021,OTFS_Eq_Learning_TWC_2022,ISAC_OTFS_JSAC_2022} and to prevent ISI between consecutive symbols in OFDM-based (multiple-CP) OTFS \cite{OTFS_CE_TSP_2019}. In Fig.~\ref{fig_b_ISI_all}, we provide an illustrative example of how the unambiguous range is increased via ISI exploitation.

Similarly to ISI, ICI can be exploited to increase the maximum detectable unambiguous Doppler by a factor of $N$ \cite{MIMO_OFDM_ICI_JSTSP_2021}. In addition to the standard \textit{slow-time} steering vector $\cc_M(\nu)$, Doppler-dependent phase rotations can also be captured by the \textit{fast-time} steering vector $\ccici(\nu)$, which allows sampling the entire time window $M T$ with an interval of $T/N$. Therefore, the unambiguous Doppler values with and without ICI can be expressed, respectively, as
\begin{align}\label{eq_ici_exp}
    \numaxici =\frac{N}{T}, ~ \numax =  \frac{1}{T} ~.
\end{align}



\section{Detection and Estimation with MIMO-OTFS Radar}\label{sec_otfs_radar}
We next design a GLRT based detection and estimation scheme for MIMO-OTFS radar that operates on the observations in \eqref{eq_obs_mimo}. The proposed detector/estimator inherently leverages the ISI/ICI exploitation capability introduced by the novel OTFS signal formulation, as discussed in Sec.~\ref{sec_isi_ici}. We note that the communication receiver processing is outside the scope of this paper, assuming that standard OTFS receive operations are performed \cite{OTFS_CE_TSP_2019,OTFS_SBL_TWC_2022,OTFS_Eq_Learning_TWC_2022}.


\subsection{GLRT for Detection/Estimation in MIMO-OTFS Radar}\label{sec_grlt}
Given the transmit signal $\boldS$ in \eqref{eq_s_tx}, the problem of interest for MIMO-OTFS radar sensing is to detect the presence of multiple targets and estimate their parameters, i.e., gains, delays, Dopplers and angles $\{(\alpha_k, \tau_k, \nu_k, \theta_k)\}_{k=0}^{K-1}$ from the observation in \eqref{eq_obs_mimo}. Unlike most of the existing works in the OTFS radar literature, e.g., \cite{Gaudio_MIMO_OTFS_Hybrid,otfs_radar_2019,OTFS_RadCom_TWC_2020,MIMO_OTFS_Radar_2020,beamspaceMIMO_OTFS_2022}, where estimator design is based on the received symbols in the DD domain, we perform detection/estimation directly using \textit{time domain} observations at multiple RX antennas without transforming them into \textit{DD domain}. As an improvement over previous OTFS radar studies \cite{Gaudio_MIMO_OTFS_Hybrid,otfs_radar_2019,OTFS_RadCom_TWC_2020,MIMO_OTFS_Radar_2020,beamspaceMIMO_OTFS_2022,OTFS_IOT_2021,ISAC_OTFS_JSAC_2022}, this approach enables exploiting the ISI and ICI effects to surpass the standard ambiguity limits in range and velocity estimation.

The hypothesis testing problem to test the presence of a single target in \eqref{eq_obs_mimo} can be expressed as
\begin{align}\label{eq_hypo}
    \boldY = \begin{cases}
	\boldZ,&~~ {\rm{under~\mathcal{H}_0}}  \\
	\alpha  \,   \boldC(\nu)  \FF^H \boldB(\tau) \FF \boldS \, \atx(\theta) \arx^T(\theta)   + \boldZ ,&~~ {\rm{under~\mathcal{H}_1}} 
	\end{cases} ~,
\end{align}
where the hypotheses $\mathcal{H}_0$ and $\mathcal{H}_1$ represent the absence and presence of a target, respectively. To solve \eqref{eq_hypo}, we treat $\alpha$, $\tau$, $\nu$ and $\theta$ as deterministic unknown parameters and resort to the GLRT
\begin{equation}\label{eq_glrt}
    \Lambda(\yy) = \frac{ \max_{\alpha, \tau, \nu, \theta} p(\boldY \, \lvert \, \mathcal{H}_1 ; \alpha, \tau, \nu, \theta ) }{p(\boldY \, \lvert \, \mathcal{H}_0  ) } \underset{\mathcal{H}_0}{\overset{\mathcal{H}_1}{\gtrless}} \eta~,
\end{equation}
where $\eta$ is a threshold set to satisfy a given probability of false alarm. Under the assumption $\vecc{\boldW} \sim \mtCN(\boldzero, \sigma^2 \Imatrix )$, the GLRT after taking the logarithm becomes
\begin{align}\label{eq_glrt2}
    \Lambdalog(\yy) = \norm{\boldY}_F^2 -  \min_{\alpha, \tau, \nu, \theta} \norm{  \boldY - \alpha \, \boldC(\nu)  \FF^H \boldB(\tau)  \FF \boldS \, \atx(\theta) \arx^T(\theta)   }_F^2 \underset{\mathcal{H}_0}{\overset{\mathcal{H}_1}{\gtrless}} \etatilde ~,
\end{align}
where $\Lambdalog(\yy) = \log \Lambda(\yy)$ and $\etatilde = \sigma^2 \log \eta$. For fixed $\tau$, $\nu$ and $\theta$, the optimal channel gain in \eqref{eq_glrt2} is given by $\alphahat = { \tracebig{ \boldA^H(\tau, \nu, \theta) \boldY }  }/{ \norm{\boldA(\tau, \nu, \theta)}_F^2 }$, 
where $\boldA(\tau, \nu, \theta) \triangleq \boldC(\nu)  \FF^H \boldB(\tau)  \FF \boldS \, \atx(\theta) \arx^T(\theta) $. Plugging $\alphahat$ back into \eqref{eq_glrt2}, we have the detection test
\begin{equation}\label{eq_glrt_final}
    \max_{\tau, \nu, \theta} \frac{ \absbig{ \tracebig{ \boldA^H(\tau, \nu, \theta) \boldY } }^2 }{ \norm{\boldA(\tau, \nu, \theta)}_F^2 } \underset{\mathcal{H}_0}{\overset{\mathcal{H}_1}{\gtrless}} \etatilde ~.
\end{equation}
For coherent processing, computing the decision statistic in \eqref{eq_glrt_final} entails a computationally prohibitive 3-D search in the delay-Doppler-angle domain \cite{MIMO_OFDM_radar_TAES_2020}. To ease the computational burden (especially for automotive applications), we propose to first perform 2-D delay-Doppler processing by noncoherent integration across the antenna elements and then estimate the angles via 1-D spatial processing at the detected delay-Doppler locations \cite{MIMO_OFDM_radar_TAES_2020,comb_MIMO_OFDM_radar_2020}, as described next. 

\subsection{Reduced-Complexity GLRT via Noncoherent Integration}
Opening up the terms in the numerator of \eqref{eq_glrt_final}, we obtain
\begin{align} \label{eq_num_glrt}
    \absbig{ \tracebig{ \boldA^H(\tau, \nu, \theta) \boldY } }^2    
    = \absbig{ \atx^H(\theta) \boldS^H \FF^H \boldB^H(\tau) \FF \boldC^H(\nu) \boldY \arx^{\ast}(\theta)  }^2 ~.
\end{align}
Noncoherent integration in the spatial domain in \eqref{eq_num_glrt} corresponds to summing up the squared magnitudes of the elements of the spatial domain matrix $\boldS^H \FF^H \boldB^H(\tau) \FF \boldC^H(\nu) \boldY \in \complexset{\Ntx}{\Nrx}$ instead of phase-aligning with the TX and RX steering vectors $\atx(\theta)$ and $\arx(\theta)$, i.e.,
\begin{align} \label{eq_num_ev}
    \norm{\boldS^H \FF^H \boldB^H(\tau) \FF \boldC^H(\nu) \boldY}_F^2 ~.
\end{align}

Similarly, the denominator of \eqref{eq_glrt_final} can be written after straightforward algebraic manipulations as
\begin{align} \label{eq_den_ev}
    \norm{\boldA(\tau, \nu, \theta)}_F^2 =  \atx^H(\theta) \boldS^H \boldS \atx(\theta) ~.
\end{align}
Following noncoherent integration across the antenna elements, we have $ \norm{\boldS^H \boldS}_F$ for the expression in \eqref{eq_den_ev}. Combining this with \eqref{eq_num_ev}, the reduced-complexity version of the GLRT in \eqref{eq_glrt_final} is given by
\begin{equation}\label{eq_glrt_final_red}
    \max_{\tau, \nu} \norm{\boldS^H \FF^H \boldB^H(\tau) \FF \boldC^H(\nu) \boldY}_F^2 \underset{\mathcal{H}_0}{\overset{\mathcal{H}_1}{\gtrless}} \etabar  ~,
\end{equation}
where $\etabar = \etatilde \norm{\boldS^H \boldS}_F$.

\subsection{Multi-Target Detection and Angle Estimation}
To account for the presence of multiple targets, the decision statistic in \eqref{eq_glrt_final_red} can be computed over a discretized delay-Doppler region and targets are declared at those locations where there is a peak exceeding the threshold \cite[Ch.~6.2.4]{richards2005fundamentals}. Let $\{\tauhat_k, \nuhat_k\}_{k=0}^{\kbar-1}$ be the delay-Doppler values of the targets detected through the GLRT in \eqref{eq_glrt_final_red}. Then, the angle estimation for the $\thn{k}$ target can be performed by solving the maximization problem in the original GLRT formulation \eqref{eq_glrt_final} after plugging the associated delay-Doppler estimates, i.e.,
\begin{align} \label{eq_ang_spec}
    \thetahat_k = \arg \max_{\theta} \frac{ \absbig{ \atx^H(\theta) \boldS^H \FF^H \boldB^H(\tauhat_k) \FF \boldC^H(\nuhat_k) \boldY \arx^{\ast}(\theta)  }^2 }{ \atx^H(\theta) \boldS^H \boldS \atx(\theta) } ~.
\end{align}
Considering the possibility that there exist multiple targets in the same delay-Doppler bin with different angles, we search for multiple peaks in the angular spectrum of \eqref{eq_ang_spec} exceeding a threshold for a given probability of false alarm. Additionally, to prevent leakage between different targets in the angular domain, an orthogonal matching pursuit (OMP)-like simple interference subtraction procedure (e.g., \cite{OMP_mmWave_2016}) is applied for refining angle estimates. The overall algorithm to detect multiple targets and estimate their delay-Doppler-angle parameters is summarized in Algorithm~\ref{alg_glrt}. It should be noted that Algorithm~\ref{alg_glrt} is generic and can be employed for any single-CP waveform $\boldS$.

\begin{algorithm}[t]
	\caption{GLRT Based 3-D MIMO-OTFS Radar Sensing}
	\label{alg_glrt}
	\begin{algorithmic}[1]
		\STATE \textbf{Input:} MIMO-OTFS radar observations $\boldY$ in \eqref{eq_obs_mimo}, probability of false alarm $\pfa$.
		\STATE \textbf{Output:} Delay-Doppler-angle estimates $\{\tauhat_k, \nuhat_k, \thetahat_k\}_{k=0}^{K-1}$ of multiple targets.
		\STATE Compute the GLRT metric in \eqref{eq_glrt_final_red} over a delay-Doppler region.
        \STATE Detect targets in delay-Doppler domain by running a CFAR detector with the given $\pfa$.
        \STATE For each detected target with $(\tauhat_k, \nuhat_k)$, compute the angular spectrum in \eqref{eq_ang_spec}.
        \STATE Estimate angles from the spectrum by running a CFAR detector with the given $\pfa$.
	\end{algorithmic} 
	\normalsize
\end{algorithm}

\section{MIMO-OTFS Signal Design for ISAC}\label{sec_track}
In this section, we propose signal design strategies (i.e., the design of the DD windows $\{\boldW_i\}_{i=1}^{\Ntx}$) to be employed in the search and track modes of the MIMO-OTFS ISAC system, mentioned in Sec.~\ref{sec_dd_window}. We begin by showing the orthogonality of transmit waveforms under the DD multiplexing strategy and the respective virtual array structure in the search mode. Then, we derive radar and communication performance metrics for the track mode, formulate the OTFS ISAC trade-off problem to optimize the DD windows and propose an algorithm based on DD-domain water-filling and Rayleigh quotient maximization to solve it.

\subsection{Search Mode: Orthogonality of Transmit Waveforms via Delay-Doppler Multiplexing}\label{sec_search_DD_mult}
In search mode, the proposed DD multiplexing strategy in \eqref{eq_dd_mult} leads to mutually orthogonal transmit waveforms, as shown in the following lemma. 
\begin{lemma}\label{lemma_dd}
   For rectangular pulse-shapes \cite{raviteja2018practical,ISAC_OTFS_JSAC_2022}, i.e., $\gtxmat = \Imatrix$, the transmit waveform matrix $\boldS$ in \eqref{eq_s_tx} with the DD multiplexing in \eqref{eq_dd_mult} satisfies
   \begin{align}\label{eq_shs}
    \boldS^H \boldS = \diag{ P_1, \ldots, P_{\Ntx} }  ~,
\end{align} 
where $P_i \triangleq \norm{\boldXdd_i}_F^2$ is the transmit power of the $\thn{i}$ antenna.
\end{lemma}
\begin{proof}
    Please see App.~\ref{app_lemma_dd}.
\end{proof}

Since \eqref{eq_shs} provides omnidirectional transmission, no signal design is needed in search mode to optimize ISAC trade-offs and we randomly select orthogonal DD windows according to \eqref{eq_dd_mult}. To observe the virtual array structure in the OTFS radar observations \eqref{eq_obs_mimo}, enabled by the orthogonality property in \eqref{eq_shs}, we consider a single-target scenario without noise, leading to
\begin{align}\label{eq_obs_mimo_single}
    \boldYs &= \alpha  \,   \boldC(\nu)  \FF^H \boldB(\tau) \FF \boldS \, \atx(\theta) \arx^T(\theta) ~.
\end{align}
Performing correlation of \eqref{eq_obs_mimo_single} with the matched filter tuned to the delay-Doppler pair $(\tau, \nu)$, i.e.,
\begin{align}\label{eq_mf_dd}
    \boldGamma(\tau, \nu) = \boldC(\nu)  \FF^H \boldB(\tau) \FF \boldS \big(\boldS^H \boldS\big)^{-1} \in \complexset{NM}{\Ntx} ~,
\end{align}
yields the spatial domain observations 
\begin{align}\nonumber
     \boldQ &= \boldGamma^H(\tau, \nu) \boldYs = \alpha  \, \big(\boldS^H \boldS\big)^{-1} \boldS^H \FF^H \boldB^H(\tau) \FF \boldC^H(\nu) 
      \boldC(\nu)  \FF^H \boldB(\tau) \FF \boldS \,  \atx(\theta) \arx^T(\theta) ~, \\ \label{eq_y_mf_dd}
     &= \alpha \, \atx(\theta) \arx^T(\theta) \in \complexset{\Ntx}{\Nrx} ~.
\end{align}
In vector form, we have
\begin{align} \label{eq_vecz_virt}
    \vecc{\boldQ} = \alpha \, \arx(\theta) \otimes \atx(\theta) \in \complexset{\Ntx \Nrx}{1}  ~,
\end{align}
where $\arx(\theta) \otimes \atx(\theta)$ represents the steering vector of a virtual array of $\Ntx \Nrx$ elements \cite{phasedMIMO_radar_TSP_2010}. Clearly, \eqref{eq_vecz_virt} verifies that the proposed DD multiplexing in \eqref{eq_dd_mult}, which assigns  non-overlapping DD bins to TX antennas, creates a virtual array for MIMO-OTFS radar, leading to improved angular resolution.

\subsection{Radar Metric for Track Mode}


According to \eqref{eq_wwi_phased}, our optimization variables of interest in track mode are the TX beamformer $\betab \triangleq [\beta_1 \ \cdots \ \beta_{\Ntx}]^T \in \complexset{\Ntx}{1}$ and the DD amplitudes $\pp \in \realset{NM}{1}$. We adopt the integrated SNR over a given delay-Doppler-angle region of interest $\Rs = \Rtau \times \Rnu \times \Rtheta$ as the radar metric: $ \snrrad(\pp, \betab) = \int \int \int_{\Rs} \snrrad(\pp, \betab; \tau, \nu, \theta ) \d \tau \d \nu \d \theta$, 
where
\begin{align} \label{eq_snrrad}
    \snrrad(\pp, \betab; \tau, \nu, \theta )  = \frac{\abs{\alpha}^2 \abs{ \betab^T \atx(\theta)}^2}{\sigma^2}
\end{align}
represents the SNR for a target located at $(\tau, \nu, \theta)$ with channel gain $\alpha$ (see App.~\ref{app_snr} for details). It is evident from \eqref{eq_snrrad} that the SNR depends neither on the DD window/power allocation $\pp$ nor on the delay-Doppler $(\tau, \nu)$ of the target. Using \eqref{eq_snrrad}, the SNR metric is given by
\begin{align} \label{eq_snrrad2}
    \snrrad(\betab)  = \betab^T \DDrad \betab^{\ast} ~,
\end{align}
where $\DDrad = \frac{1}{\sigma^2} \sum_{k=0}^{K-1}  \abs{\alphahat_k}^2   \atx(\thetahat_k) \atx^H(\thetahat_k)$, 
with $\alphahat_k$ and $\thetahat_k$ denoting the estimated gain and angle parameters of the targets.



\subsection{Communication Metric for Track Mode}
We now derive the communication metric to be employed in the track mode. The LMMSE estimate of the transmit DD symbols based on the received DD symbols in \eqref{eq_yycom_dd_comp} is given by \cite{windowDesign_OTFS_TCOM_2021} $ \xxddhat = \boldHdd^H \big( \boldHdd \boldHdd^H + \sigma^2 \Imatrix  \big)^{-1} \yycomdd$, 
with the corresponding covariance matrix
\begin{align} \label{eq_lmmse_cov}
    \rlmmmse &\triangleq \Eee \big\{ (\xxddhat - \xxdd) (\xxddhat - \xxdd)^H \big\} = \Big( \Imatrix + \frac{1}{\sigma^2} \boldHdd^H \boldHdd  \Big)^{-1} \in \complexset{NM}{NM} ~.
\end{align}
The LMMSE covariance matrix in \eqref{eq_lmmse_cov} can be used to provide an expression for an achievable rate of the OTFS communication channel characterizing the DD domain input-output relation in \eqref{eq_yycom_dd_comp} \cite{LMMSE_capacity,LMMSE_TSP_2017,imperfectCSI_MIMO_2010,OFDM_DFRC_TSP_2021}:
\begin{align} \label{eq_cap_lmmse}
    R(\pp, \betab) = - \log \det \rlmmmse(\pp, \betab) ~.
\end{align}
In the following lemma, we derive an expression for $\rlmmmse$ in terms of the optimization variables $\pp$ and $\betab$.

\begin{lemma}\label{lemma_lmmse}
    In track mode where the DD windows are given in \eqref{eq_wwi_phased}, the LMMSE covariance matrix in \eqref{eq_lmmse_cov} can be expressed as a function of the TX beamformer $\betab$ and the DD domain amplitudes $\pp$ as
    \begin{align} \label{eq_rlmmse_lem}
        \rlmmmse(\pp, \betab) = \big( \Imatrix + (\pp \pp^T) \odot \boldG \big)^{-1}~,
    \end{align}
    where
    \begin{align} \label{eq_gmat}
    \boldG &=  \frac{1}{\sigma^2} (\FF_M \otimes \Imatrix_N) \boldHt^H 
        \boldHt (\FF_M^H \otimes \Imatrix_N) ~, 
        \\ \label{eq_ht_mat}
   \boldHt &= \sum_{k=0}^{\Ktilde-1}  \alphat_k  \betab^T \atx(\thetat_k)   \boldC(\nut_k)  \FF^H \boldB(\taut_k) \FF   ~.
    \end{align}
\end{lemma}
\begin{proof}
    Please see App.~\ref{app_proof_lmmse}. 
\end{proof}

The following lemma presents an approximation of \eqref{eq_cap_lmmse} that will enable formulating the ISAC trade-off signal design problem for $\betab$ without the knowledge of $\pp$. 
\begin{lemma}\label{lemma_approx_beta}
    Under the condition that $M$ is small or $\deltaf$ is large or channel Doppler spread is small, the achievable rate \eqref{eq_cap_lmmse} can be approximated as
    \begin{align}  \label{eq_c_approx}
        R(\pp, \betab) &\approx 
        \sum_{i=0}^{NM-1} \log\big( 1 + q_i \betab^T \DDcom \conj{\betab} \big) ~,
    \end{align}
    where  $\qq = \pp \odot \pp = [q_0 \ \cdots \ q_{NM-1}]^T$ and $\DDcom = \frac{1}{\sigma^2} \sum_{k=0}^{\Ktilde-1}  \abs{\alphat_k}^2   \atx(\thetat_k) \atx^H(\thetat_k)$.
\end{lemma}
\begin{proof}
    Please see App.~\ref{app_diag}.
\end{proof}


\subsection{MIMO-OTFS ISAC Trade-off Signal Design}\label{sec_des_track}
Based on Lemma~\ref{lemma_approx_beta}, optimization over $\betab$ to maximize the metric in \eqref{eq_c_approx} can be carried out independently from $\pp$. Hence, we first optimize the spatial domain ($\betab$) degrees of freedom to achieve the best OTFS ISAC trade-off and then optimize the DD domain ($\pp$) degrees of freedom given the optimal ISAC beamformer.

\subsubsection{Optimize $\betab$}
We employ the communication metric in \eqref{eq_c_approx} and the radar metric in \eqref{eq_snrrad2} to formulate the ISAC trade-off optimization problem for the TX beamformer $\betab$ as follows: 
\begin{subequations} \label{eq_problem_tradeoff2}
	\begin{align} \label{eq_problem_tradeoff_obj2}
	\betabopt = \arg \max_{\betab} &~~ \rho \,\betab^T \DDrad \betab^{\ast} + (1-\rho) \, \betab^T \DDcom \betab^{\ast}
	  \\ \label{eq_problem_tradeoff_obj_cons2}
	\mathrm{s.t.} &~~ \norm{\betab}_2^2 \leq 1 ~,
	\end{align} 
\end{subequations}
where the ISAC weight $0 \leq \rho \leq 1$ governs the trade-off between radar and communications \cite{DFRC_Waveform_Design}. The problem \eqref{eq_problem_tradeoff2} represents a Rayleigh quotient maximization problem\footnote{Specifically, the problem is defined as $\max_{\betab} \betab^T \boldD_{\rho} \betab^{\ast}  / \norm{\betab}_2^2 $, where $\boldD_{\rho} = \rho \DDrad + (1-\rho) \DDcom$.}, whose optimal solution $\betabopt$ is given by the conjugate of the dominant eigenvector of $\boldD_{\rho}$.




\subsubsection{Optimize $\pp$ for Given $\betab$}
Since the radar metric in \eqref{eq_snrrad2} does not depend on $\pp$, we optimize only the communication metric to determine the optimal $\pp$. Given the optimal ISAC beamformer $\betabopt$ from \eqref{eq_problem_tradeoff2}, we revert back to the original objective \eqref{eq_cap_lmmse} and derive another approximation which will be employed to optimize $\pp$ (or, equivalently $\qq$) in a tractable manner, as specified in the following lemma. 
\begin{lemma}\label{lemma_lb}
    Assuming that $\boldG$ in \eqref{eq_gmat} has small off-diagonal elements (justified in the proof of Lemma~\ref{lemma_approx_beta}), the objective \eqref{eq_cap_lmmse} can be approximated as $ R(\pp, \betab) \approx \sum_{i=0}^{NM-1} \log (1 + q_i g_i)$, 
    where $\ggb = [g_0 \ \cdots \ g_{NM-1}]^T = \diag{\boldG}$.
\end{lemma}
\begin{proof}
    Please see App.~\ref{app_proof_lb}.
\end{proof}

Based on Lemma~\ref{lemma_lb}, we propose to optimize $\qq$ as follows:
\begin{subequations} \label{eq_problem_tradeoff_subpp2}
	\begin{align} \label{eq_problem_tradeoff_subpp_obj2}
	\qqopt =  \arg \max_{\qq} &~~ \sum_{i=0}^{NM-1} \log (1 + q_i g_i)
	  \\ \label{eq_problem_tradeoff_subpp_cons2}
	\mathrm{s.t.} &~~ \boldone^T \qq \leq NM, \, \qq \succeq \boldzero ~,
	\end{align} 
\end{subequations}
where $g_i$'s are computed in \eqref{eq_gmat} and \eqref{eq_ht_mat} by plugging $\betab = \betabopt$. The problem \eqref{eq_problem_tradeoff_subpp2} represents a classical rate maximization problem whose optimal solution is given by \textit{DD-domain water-filling} with respect to the diagonals $\ggb$ of the DD correlation matrix $\boldG$ \cite[Ch.~(4.4.1)]{goldsmith2005wireless}.

\subsubsection{Overall Algorithm for ISAC Trade-off Signal Design}
Summarizing the overall algorithm, we first solve \eqref{eq_problem_tradeoff2} using the OTFS radar and communication channel parameters (angles and gains), which appear in the matrices $\DDrad$ and $\DDcom$. Then, we solve \eqref{eq_problem_tradeoff_subpp2} by plugging the resulting $\betabopt$ into \eqref{eq_gmat} and \eqref{eq_ht_mat}, and obtain $\ppopt = \sqrt{\qqopt}$. The overall algorithm to design the OTFS ISAC trade-off signal is provided in Algorithm~\ref{alg_tradeoff}.

\begin{algorithm}[t]
	\caption{Algorithm to Design OTFS ISAC Trade-off Signal in Spatial and Delay-Doppler Domains}
	\label{alg_tradeoff}
	\begin{algorithmic}[1]
		\STATE \textbf{Input:} Radar target parameters $\{\alphahat_k,\thetahat_k\}_{k=0}^{K-1}$, communication channel parameters $\{\alphat_k,\taut_k,\nut_k,\thetat_k\}_{k=0}^{\Ktilde-1}$, ISAC trade-off parameter $\rho$.
		\STATE \textbf{Output:} ISAC TX beamformer $\betabopt$, DD amplitudes $\ppopt$.
		\STATE Solve \eqref{eq_problem_tradeoff2} via Rayleigh quotient maximization to obtain $\betabopt$.
		\STATE Using $\betabopt$, solve \eqref{eq_problem_tradeoff_subpp2} via DD-domain water-filling.
	\end{algorithmic}
	\normalsize
\end{algorithm}

\section{Numerical Results}
In this section, we assess the performance of radar sensing in Alg.~\ref{alg_glrt} and investigate OTFS ISAC trade-offs through Alg.~\ref{alg_tradeoff}. The number of antennas at the ISAC transceiver is taken as $\Ntx = \Nrx = 8$. Here, the TX and the radar RX are equipped with uniform linear arrays (ULAs) with $\lambda/2$ and $\Ntx\lambda/2$ elements spacings, respectively, to enable constructing a virtual ULA of $\Ntx \Nrx = 64$ elements in search mode via orthogonal transmission, as specified in \eqref{eq_dd_mult}. For a target/path with channel gain $\alpha$, we define the signal-to-noise ratio (SNR) as $\snr = \abs{\alpha}^2/\sigma^2$. In addition, a rectangular pulse shaping waveform \cite{OTFS_IOT_2021,ISAC_OTFS_JSAC_2022} is used for $\gtx(t)$ in \eqref{eq_smt}.


\subsection{Performance of MIMO-OTFS Radar Sensing}

\begin{table}
\caption{OTFS Parameter Sets for Radar Simulations}
\centering
    \begin{tabular}{|l|l|l|}
        \hline
        \textbf{Parameter} & \textbf{ISI-dominant Regime} & \textbf{ICI-dominant Regime} \\ 
         \hline
        Carrier Frequency, $\fc$ & $28 \, \rm{GHz}$ & $28 \, \rm{GHz}$ \\ \hline
        Subcarrier Spacing, $\deltaf$ & $480 \, \rm{kHz}$ & $30 \, \rm{kHz}$  \\ \hline
        Number of Subcarriers, $N$ & $64$ & $1024$ \\ \hline
        Total Bandwidth, $B$ & $30.72 \, \rm{MHz}$ & $30.72 \, \rm{MHz}$ \\ \hline
        Symbol Duration, $T$ & $2.08 \, \rm{\mu s}$ & $33.33 \, \rm{\mu s}$ \\ \hline 
        Cyclic Prefix Duration, $\Tcp$ & $12.5 \, \rm{\mu s}$ & $12.5 \, \rm{\mu s}$ \\ \hline 
        Number of Symbols, $M$ & $128$ & $8$ \\ \hline
                
        Maximum Range, $\Rmax$ (\textit{Standard}) & $312.5 \, \rm{m}$ & $5000 \, \rm{m}$ \\ 
        \hline
        Maximum Range, $\rmaxisi$ (\textit{ISI-embracing}) & $1875 \, \rm{m}$ & no practical limit 
        \\ \hline
        Maximum Velocity, $\vmax$ (\textit{Standard}) & $\pm 1285.7 \, \rm{m/s}$ & $\pm 80.35 \, \rm{m/s}$ \\ 
         \hline
        Maximum Velocity, $\vmaxici $ (\textit{ICI-embracing}) & no practical limit & no practical limit \\
         \hline
    \end{tabular}
    \label{tab_parameters}
    \vspace{-0.1in}
\end{table}


To evaluate the performance of the MIMO-OTFS radar sensing algorithm in Alg.~\ref{alg_glrt}, we assume \textit{search mode} operation with DD multiplexing in \eqref{eq_dd_mult} and  consider two different parameter sets at mmWave, as shown in Table~\ref{tab_parameters}, for illustrating the results in both ISI-dominant (i.e., high $\deltaf$) and ICI-dominant (i.e., small $\deltaf$) operation regimes. In the ISI-dominant regime, maximum range is the limiting factor for radar detection performance, while in the ICI-dominant regime, radar performance is mainly limited by maximum velocity. The MIMO-OTFS radar observations are generated using \eqref{eq_yt} instead of the derived compact model in \eqref{eq_obs_mimo} to provide an implicit verification of the transition from \eqref{eq_yt} to \eqref{eq_obs_mimo}. The transmit symbols $\boldXdd$ in \eqref{eq_xi_dd} are chosen randomly from a 64-QAM alphabet. As a benchmark, we consider a standard 2-D FFT based processing \cite{RadCom_Proc_IEEE_2011,OFDM_Radar_Phd_2014,ofdm_radar_correlation_TAES_2020} traditionally employed in MIMO-OFDM radar \cite{comb_MIMO_OFDM_radar_2020}, where delay-Doppler images from all spatial channels are noncoherently integrated to perform detection in the delay-Doppler domain, followed by angle estimation, as done in Alg.~\ref{alg_glrt}. 
To evaluate detection performances for both Alg.~\ref{alg_glrt} and the FFT benchmark, a cell-averaging CFAR detector is employed with the probability of false alarm $\pfa = 10^{-4}$ to declare targets in the delay-Doppler domain. 

\subsubsection{ISI-Dominant Regime}
\begin{figure}
	\centering
	\includegraphics[width=0.7\linewidth]{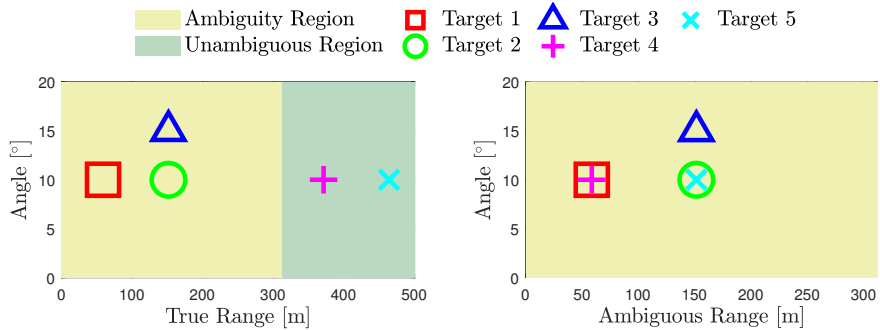}
	\caption{Range-angle scenario for MIMO-OTFS sensing in the ISI-dominant regime, where target SNRs are given by $\{ 20, 15, 5, 25, 10 \} \, \rm{dB}$, respectively. All targets have the same velocity $20 \, \rm{m/s}$. Conventional FFT-based algorithms can detect at most three targets (right), while Alg.~\ref{alg_glrt} can go beyond the standard maximum range through ISI exploitation and detect all targets (left).}
	\label{fig_scenario_isi_dominant}
	\vspace{-0.1in}
\end{figure}

In the ISI-dominant regime, we consider a scenario with five targets with the same velocity ($20\, \rm{m/s}$), but with different ranges and angles, as shown in Fig.~\ref{fig_scenario_isi_dominant}. As shown in Fig.~\ref{fig_scenario_isi_dominant} (right), the standard FFT processing \cite{RadCom_Proc_IEEE_2011,OFDM_Radar_Phd_2014,comb_MIMO_OFDM_radar_2020} can detect at most three targets since Target~4 and Target~5 fall into the same range-velocity-angle bin as Target~1 and Target~2, respectively. 

\begin{figure}
        \begin{center}
        \subfigure[]{
			 \label{fig_rp_isi_1}
			 \includegraphics[width=0.4\textwidth]{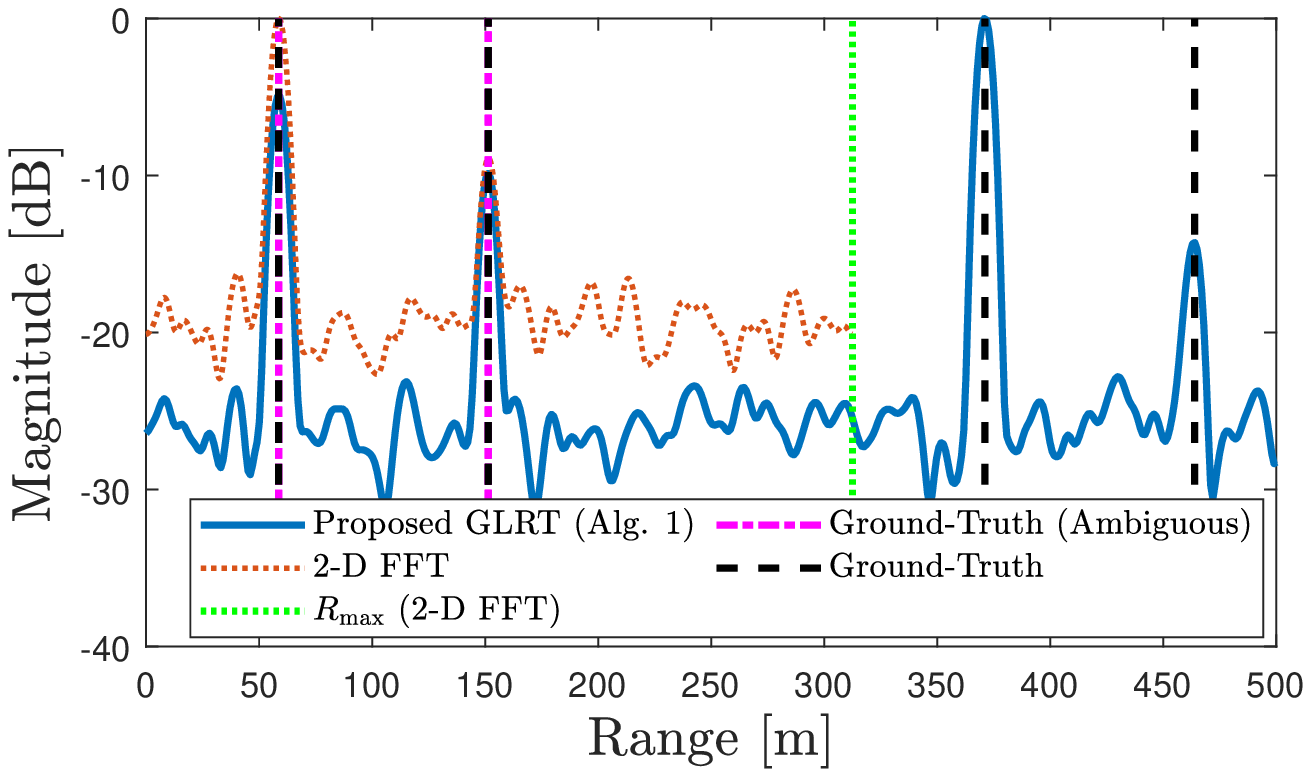}
		}
        \subfigure[]{
			 \label{fig_rp_isi_2}
			 \includegraphics[width=0.4\textwidth]{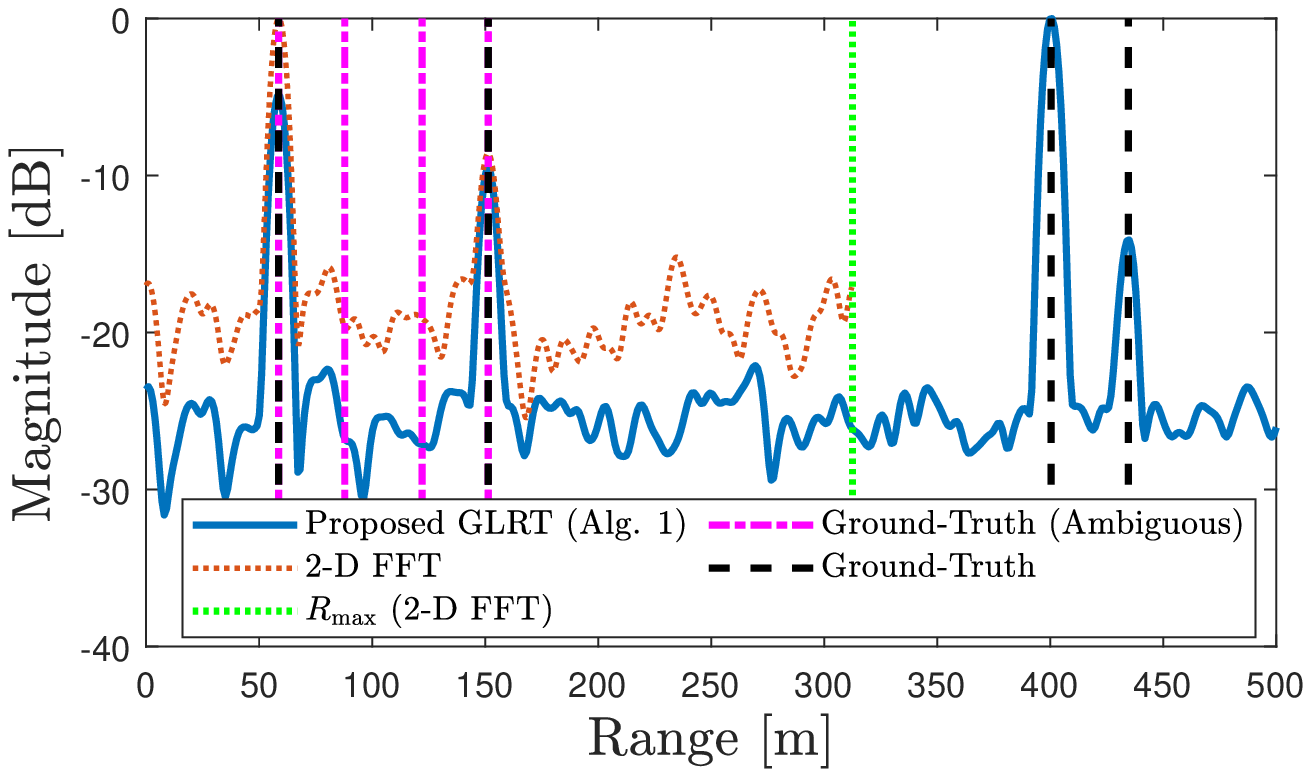}
		}
		
		\end{center}
		\vspace{-0.2in}
        \caption{ISI-dominant regime: Range profile at $v = 20 \, \rm{m/s}$ obtained by the different methods \subref{fig_rp_isi_1} for the scenario in Fig.~\ref{fig_scenario_isi_dominant}, and \subref{fig_rp_isi_2} for a modified version of the scenario in Fig.~\ref{fig_scenario_isi_dominant}, where Target~4 moved $30 \, \rm{m}$ further and Target~5 moved $30 \, \rm{m}$ closer.}  
        \label{fig_rp_isi}
        \vspace{-0.12in}
\end{figure}

Fig.~\ref{fig_rp_isi} shows an instance of the range profiles\footnote{The range profile associated with Alg.~\ref{alg_glrt} is obtained by plotting the decision statistic in \eqref{eq_glrt_final_red} for a fixed Doppler $\nu$ over an interval of delay values $\tau$. For the 2-D FFT method, range profile corresponds to the range slice of the 2-D FFT output, taken from a certain Doppler $\nu$.} of the considered methods after noncoherent integration along the spatial channels in two different scenarios. It is observed that by virtue of \textit{ISI exploitation}, the proposed GLRT detector in \eqref{eq_glrt_final_red} can detect four target ranges separately (Target~2 and Target~3 will be resolved later in angle domain via coherent spatial processing in \eqref{eq_ang_spec}) by increasing the maximum range by a factor of $6$ (see \eqref{eq_isi_exp} and Table~\ref{tab_parameters}), whereas the 2-D FFT yields peaks only at the locations of Target~1 and Target~2. In addition, the GLRT detector achieves lower side-lobe levels than the 2-D FFT method by taking into account the ISI in detector design, which enables compensating for its effect on the range profile. Moreover, even when Target~4 and Target~5 are displaced in Fig.~\ref{fig_rp_isi_2} so that four targets are resolvable in the ambiguity region, the 2-D FFT can only detect Target~1 due to the strong ISI effect, while all the target peaks are clearly visible in the range profile of the GLRT detector. Therefore, the proposed approach can simultaneously mitigate ISI to have low side-lobes and embrace the information conveyed by ISI to detect targets beyond the standard maximum range limit $\Rmax$.

\begin{figure}%
\centering
\subfigure[]{%
\label{fig_pd_snr_isi}
\includegraphics[height=1.5in]{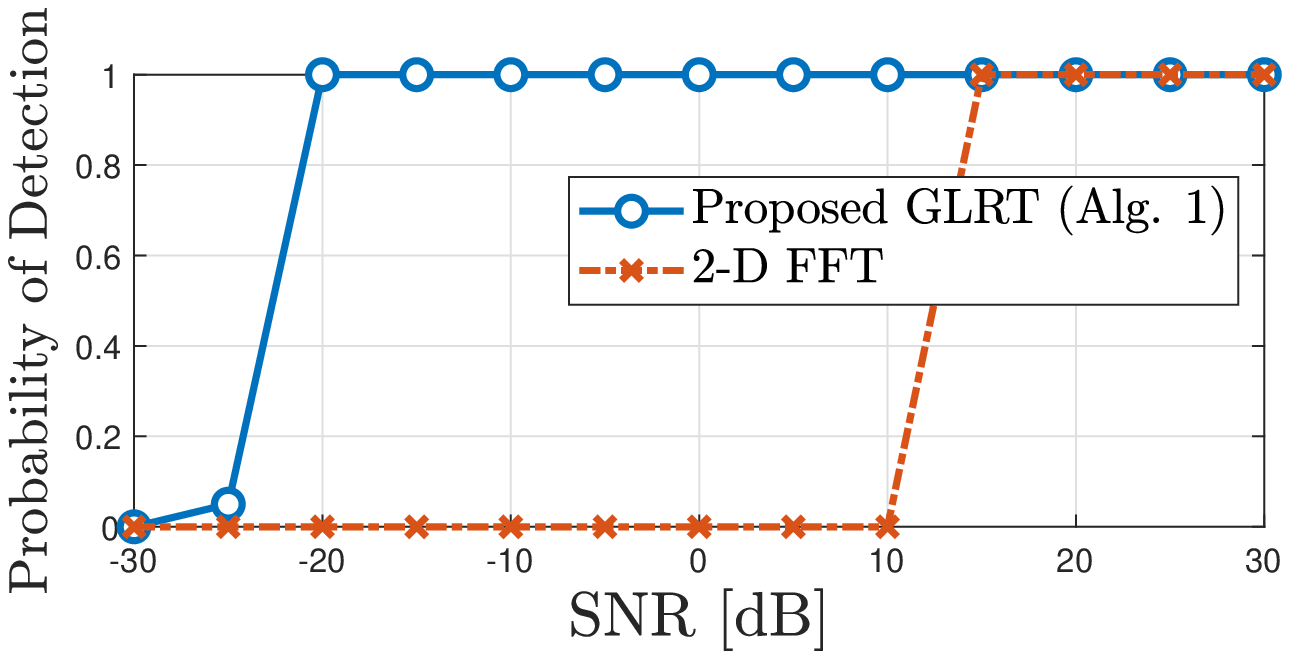}}%
\qquad
\subfigure[]{%
\label{fig_range_rmse_isi}
\includegraphics[height=1.5in]{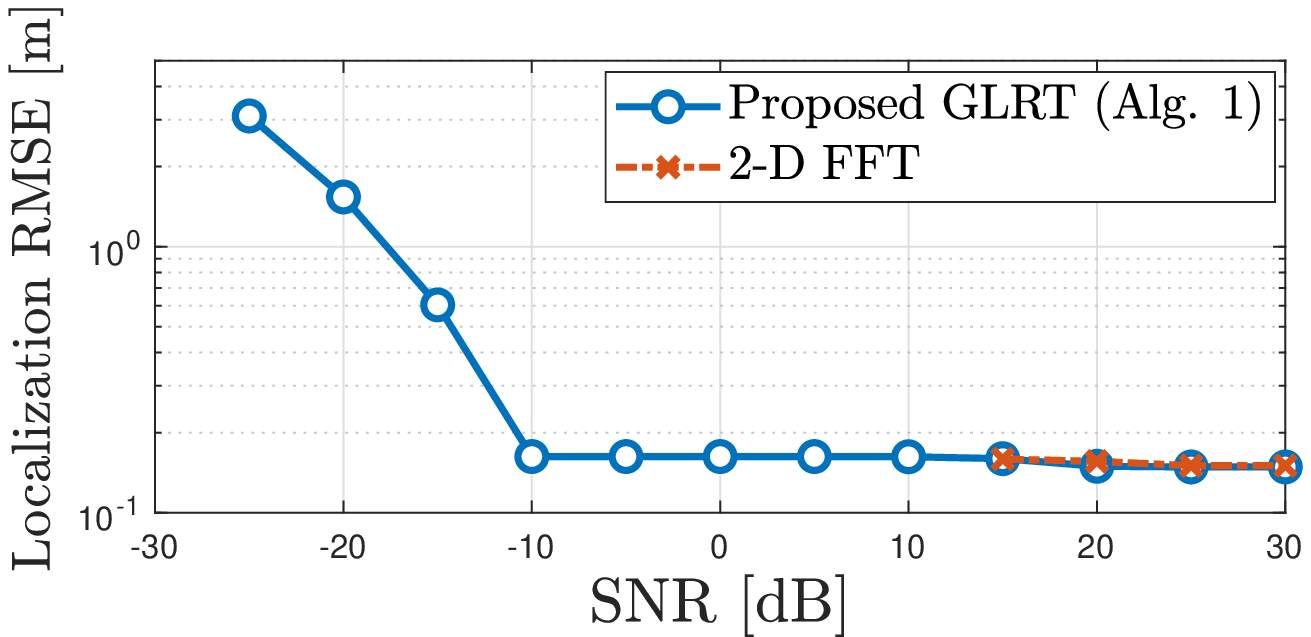}}%
 \vspace{-0.1in}
\caption{Detection and estimation performances of the proposed MIMO-OTFS radar processing algorithm (Alg.~\ref{alg_glrt}) and the FFT benchmark with respect to SNR in the ISI-dominant regime. \subref{fig_pd_snr_isi} Probability of detection, 
        and \subref{fig_range_rmse_isi} localization RMSE.}
        \label{fig_isi_det_est}
        \vspace{-0.2in}
\end{figure}

To illustrate the detection and location estimation performance of Alg.~\ref{alg_glrt}, we simulate $100$ independent Monte Carlo noise realizations and choose Target~3 as the reference target. Fig.~\ref{fig_isi_det_est} shows the probability of detection and the root mean-squared error (RMSE) of location estimates of the reference target with respect to its SNR. It is seen that the proposed detector/estimator in Alg.~\ref{alg_glrt} significantly outperforms the standard FFT method in terms of the detection performance, while they exhibit similar estimation performances (only above a certain SNR where the target is detected by the FFT method). This performance boost is accomplished through the ISI-aware modeling in \eqref{eq_obs_mimo} and the corresponding detector design in \eqref{eq_glrt_final_red}, which performs ISI compensation via the term $\boldB^H(\tau)$ and helps suppression of ISI-induced side-lobe levels, in agreement with Fig.~\ref{fig_rp_isi}. Moreover, Fig.~\ref{fig_isi_det_est} also indicates that Target~2 and Target~3 can be resolved in the angular domain using a virtual ULA of $64$ elements in search mode. We note that the GLRT detector/estimator in \eqref{eq_glrt_final_red} performs block-wise processing of the entire OTFS frame ($NM$ symbols), while the FFT method applies separate $N$- and $M$-point FFTs over frequency and time domains, respectively, which provides computational simplicity, but leads to poor radar performance.

		


\subsubsection{ICI-Dominant Regime}
\begin{figure}
	\centering
	\includegraphics[width=0.7\linewidth]{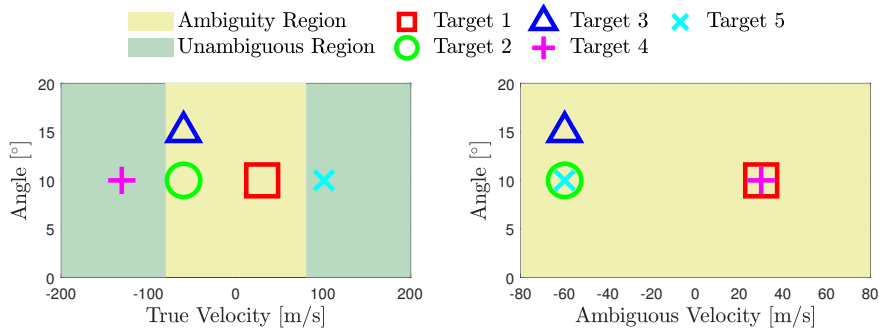}
	\caption{Velocity-angle scenario for MIMO-OTFS sensing in the ICI-dominant regime, where target SNRs are given by $\{ 20, 15, 5, 25, 10 \} \, \rm{dB}$, respectively. All targets have the same range $100 \, \rm{m}$. The proposed GLRT-based sensing algorithm can detect all targets separately via ICI exploitation (left), whereas standard FFT-based algorithms can detect only three targets due to maximum velocity limit dictated by symbol duration (right).}
	\label{fig_scenario_ici_dominant}
	\vspace{-0.2in}
\end{figure}

In the ICI-dominant regime, a scenario with five targets located at the same range ($100\, \rm{m}$), but with different velocities and angles is considered, as shown in Fig.~\ref{fig_scenario_ici_dominant}. Similar to the ISI-dominant regime, at most three targets can be detected via conventional FFT processing as the high-speed targets fall into the same bin as the (relatively) low-speed ones, as seen from Fig.~\ref{fig_scenario_ici_dominant} (right).

\begin{figure}
	\centering
	\includegraphics[width=0.45\linewidth]{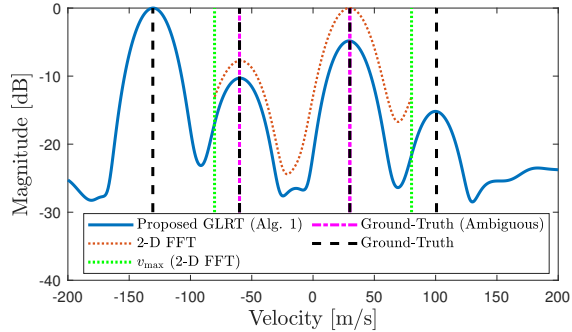}
 \vspace{-0.2in}
	\caption{ICI-dominant regime: Velocity profile at $R = 100 \, \rm{m}$ obtained by the different methods for the scenario in Fig.~\ref{fig_scenario_ici_dominant}.}
	\label{fig_vp_ici}
	\vspace{-0.2in}
\end{figure}

\begin{figure}[t]
	\centering
	\includegraphics[width=0.45\linewidth]{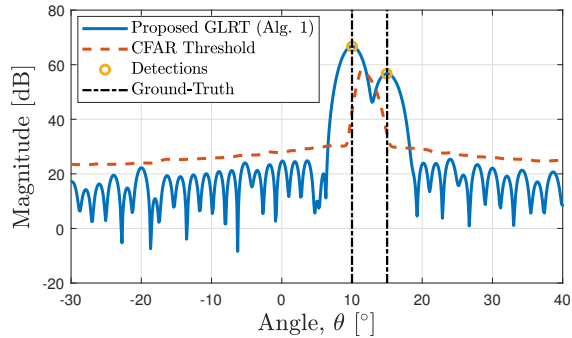}
 \vspace{-0.2in}
	\caption{Angular profile at $R = 100 \, \rm{m}$ and $v = -60 \, \rm{m/s}$, obtained by Alg.~\ref{alg_glrt} for the scenario in Fig.~\ref{fig_scenario_ici_dominant}. The DD multiplexing in \eqref{eq_dd_mult} creates a virtual ULA of $64$ elements and enables resolving Target~2 and Target~3 located $5\degree$ apart from each other using TX and RX ULAs of 8 elements each (which normally have $12\degree$ beamwidth with non-orthogonal transmission).}
	\label{fig_angular_profile_ici}
	\vspace{-0.1in}
\end{figure}

We illustrate an instance of the velocity profiles obtained by the considered methods in Fig.~\ref{fig_vp_ici}. As indicated in \eqref{eq_ici_exp} and Table~\ref{tab_parameters}, the proposed approach increases the maximum velocity by a factor of $N = 1024$, which allows detection of targets beyond the standard velocity limit $\vmax$. Hence, in the range-velocity domain, the proposed approach in Alg.~\ref{alg_glrt} can resolve four targets and detect their true (unambiguous) velocities through its \textit{ICI exploitation} capability, while the 2-D FFT method can only detect two targets. Moreover, as seen from Fig.~\ref{fig_angular_profile_ici}, Target~2 and Target~3 are resolved later in the angular domain after delay-Doppler processing (see lines 5 and 6 in Alg.~\ref{alg_glrt}). Therefore, all five targets can be resolved through ICI exploitation and via the proposed 3-D MIMO-OTFS sensing algorithm. Fig.~\ref{fig_angular_profile_ici} also corroborates the orthogonality of transmit waveforms generated according to the proposed DD multiplexing strategy, as specified in Lemma~\ref{lemma_dd}. Specifically, an RX ULA of $8$ elements has approximately $12\degree$ beamwidth, which is not sufficient to resolve Target~2 and Target~3 separated by $5\degree$. With a virtual ULA of $64$ elements (which has approximately $2\degree$ beamwidth), it becomes possible to resolve Target~2 and Target~3 via coherent spatial processing in Alg.~\ref{alg_glrt}.

\begin{figure}[t]%
\centering
\subfigure[]{%
\label{fig_pd_snr_ici}
\includegraphics[height=1.5in]{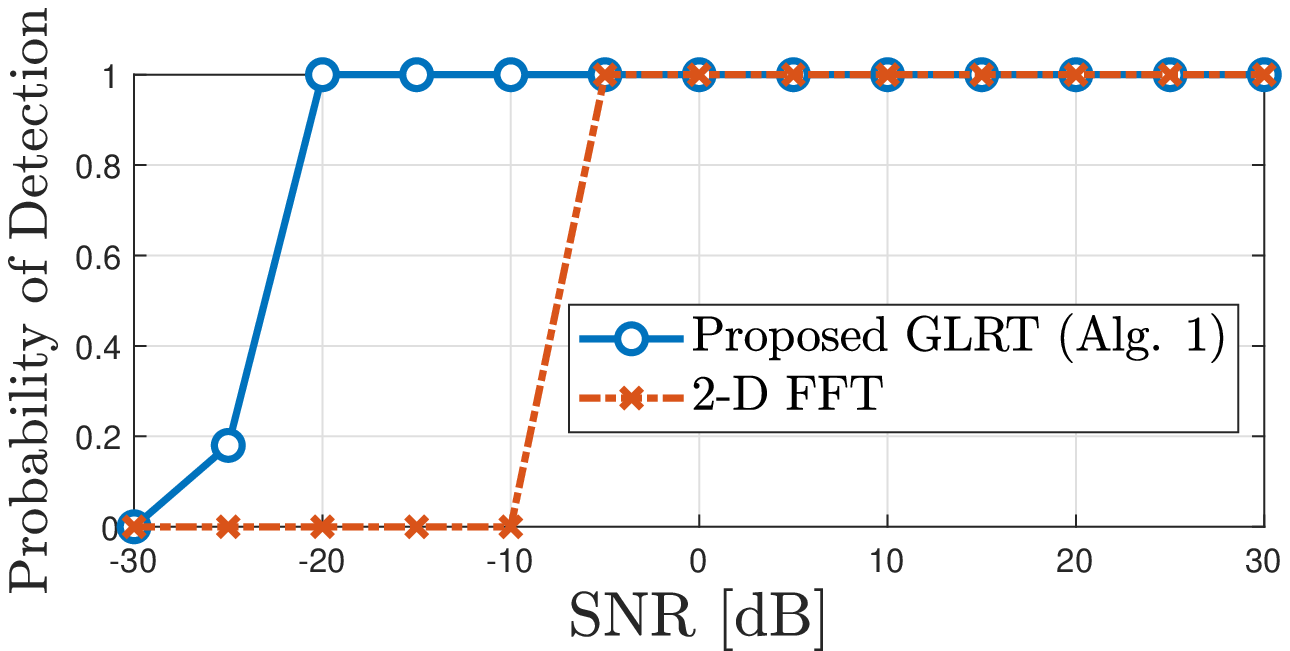}}%
\qquad
\subfigure[]{%
\label{fig_loc_rmse_ici}
\includegraphics[height=1.5in]{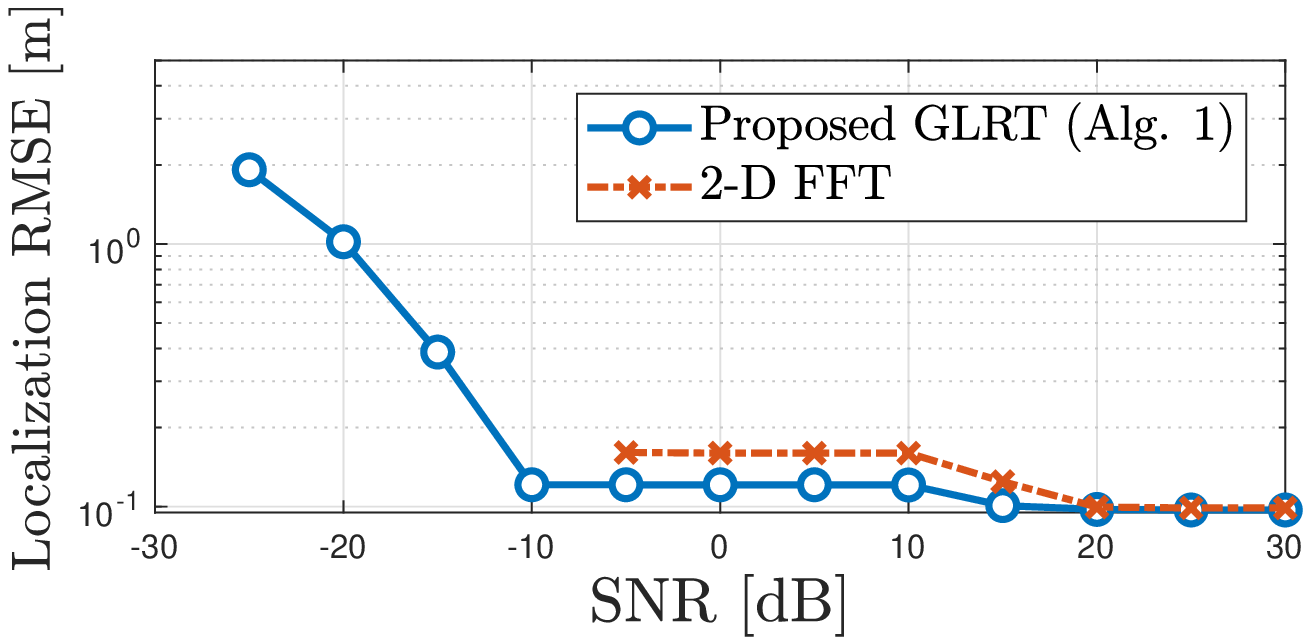}}%
 \vspace{-0.1in}
\caption{Detection and estimation performances of the proposed MIMO-OTFS radar processing algorithm (Alg.~\ref{alg_glrt}) and the FFT benchmark with respect to SNR in the ICI-dominant regime. \subref{fig_pd_snr_ici} Probability of detection, 
        and \subref{fig_loc_rmse_ici} localization RMSE.}
        \label{fig_ici_det_est}
        \vspace{-0.2in}
\end{figure}

Finally, the detection and location estimation performances in the ICI-dominant regime are shown in Fig.~\ref{fig_ici_det_est}, with respect to the SNR of the reference target, Target~3, averaged over $100$ realizations. Similar to the ISI-dominant case, the proposed approach in Alg.~\ref{alg_glrt} achieves significant performance gains over the conventional FFT method, especially in terms of  the probability of detection by explicitly accounting for the ICI effect in detector/estimator design.

		


\subsection{Evaluation of OTFS ISAC Trade-offs}
\begin{figure}
	\centering
	\includegraphics[width=0.4\linewidth]{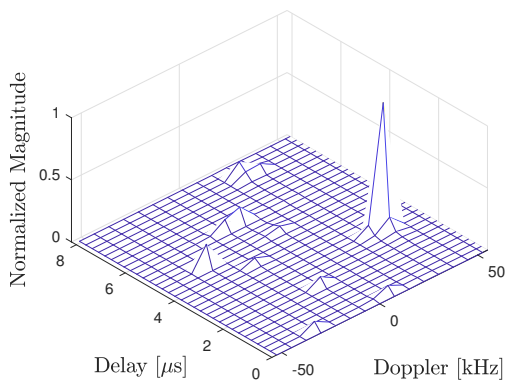}
	\caption{Impulse response of the delay-Doppler communication channel in \eqref{eq_miso_channel_comm} with the LOS-to-multipath ratio (LMR) $\Kric = 0 \, \rm{dB}$.}
	\label{fig_dd_channel}
	\vspace{-0.2in}
\end{figure}

In this part, we evaluate OTFS ISAC trade-offs in \textit{track mode}, as described in \eqref{eq_wwi_phased}, using the proposed signal design approach in Alg.~\ref{alg_tradeoff}. An OTFS system with $N = 32$, $M = 16$, $\deltaf = 120 \, \rm{kHz}$, $\fc = 28 \, \rm{GHz}$ is considered. For the communication channel in \eqref{eq_miso_channel_comm}, we consider $\Ktilde = 11$ paths and define the line-of-sight (LOS)-to-multipath ratio (LMR) parameter \cite{RIS_Beamform_JSTSP_2022} as $\Kric = \abs{\alphat_0}^2/\sum_{k=1}^{\Ktilde-1} \abs{\alphat_k}^2$ ($k=0$ denotes the LOS path) to explore ISAC trade-offs under different multipath conditions. For a given $\Kric$, $\alphat_k$'s are randomly generated such that the total SNR satisfies $\sum_{k=0}^{\Ktilde-1} \abs{\alphat_k}^2 / \sigma^2 = 25 \, \rm{dB}$. To separately observe spatial and DD domain effects in isolation from each other, the path angles are set to $\thetat_k = -30 \degree ~ \forall k$, while the path delays and Dopplers are randomly generated. The resulting delay-Doppler channel with $\Kric = 0 \, \rm{dB}$ is shown in Fig.~\ref{fig_dd_channel}. For the radar channel, we adopt the same scenario as shown in Fig.~\ref{fig_scenario_isi_dominant}.

\begin{figure}
    \centering
    \begin{minipage}{0.48\textwidth}
        \centering
        \includegraphics[width=0.9\textwidth]{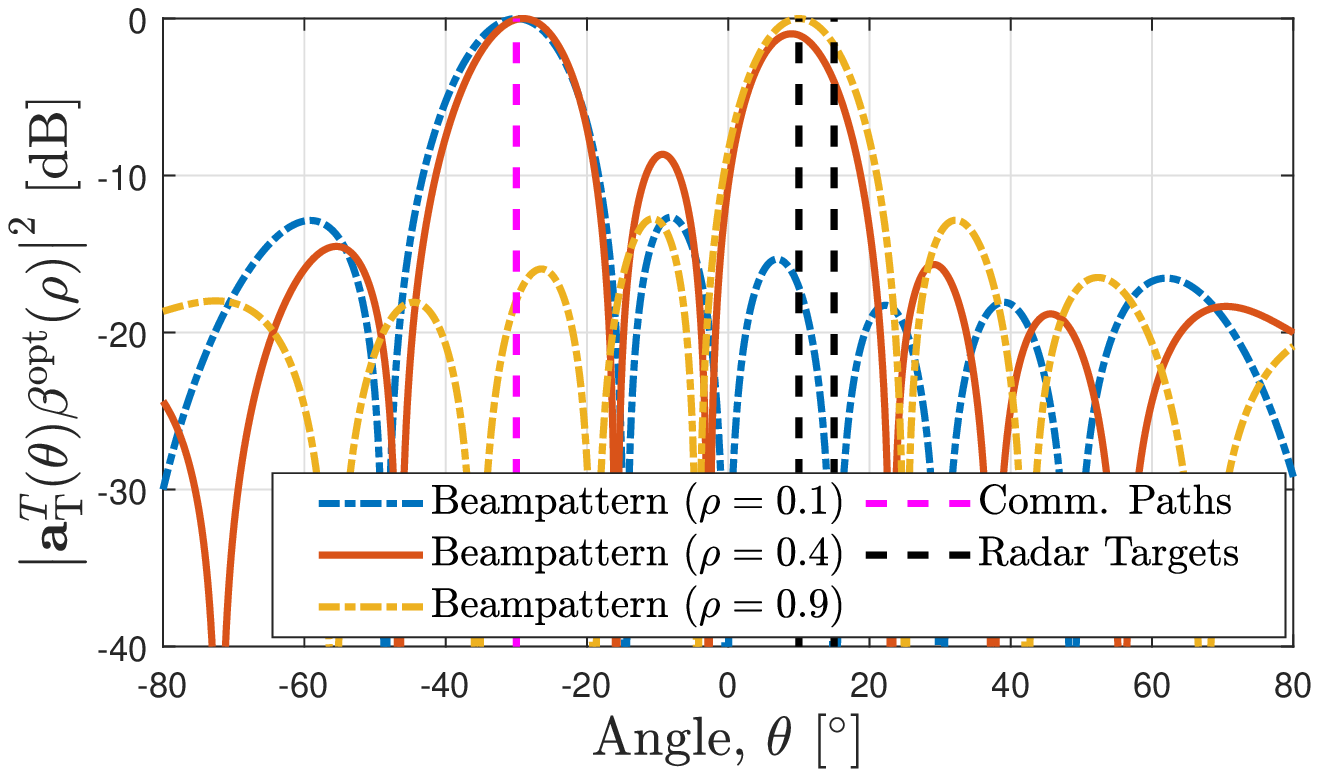} 
        \vspace{-0.2in}
        \caption{ISAC transmit beampatterns for various values of the trade-off parameter $\rho$, where $\betabopt(\rho)$ is obtained via Alg.~\ref{alg_tradeoff} and $\Kric = 0 \, \rm{dB}$.}        
        \label{fig_beampattern_tradeoff}
    \end{minipage}\hfill
    \vspace{-0.2in}
    \begin{minipage}{0.48\textwidth}
        \centering
        \includegraphics[width=0.9\textwidth]{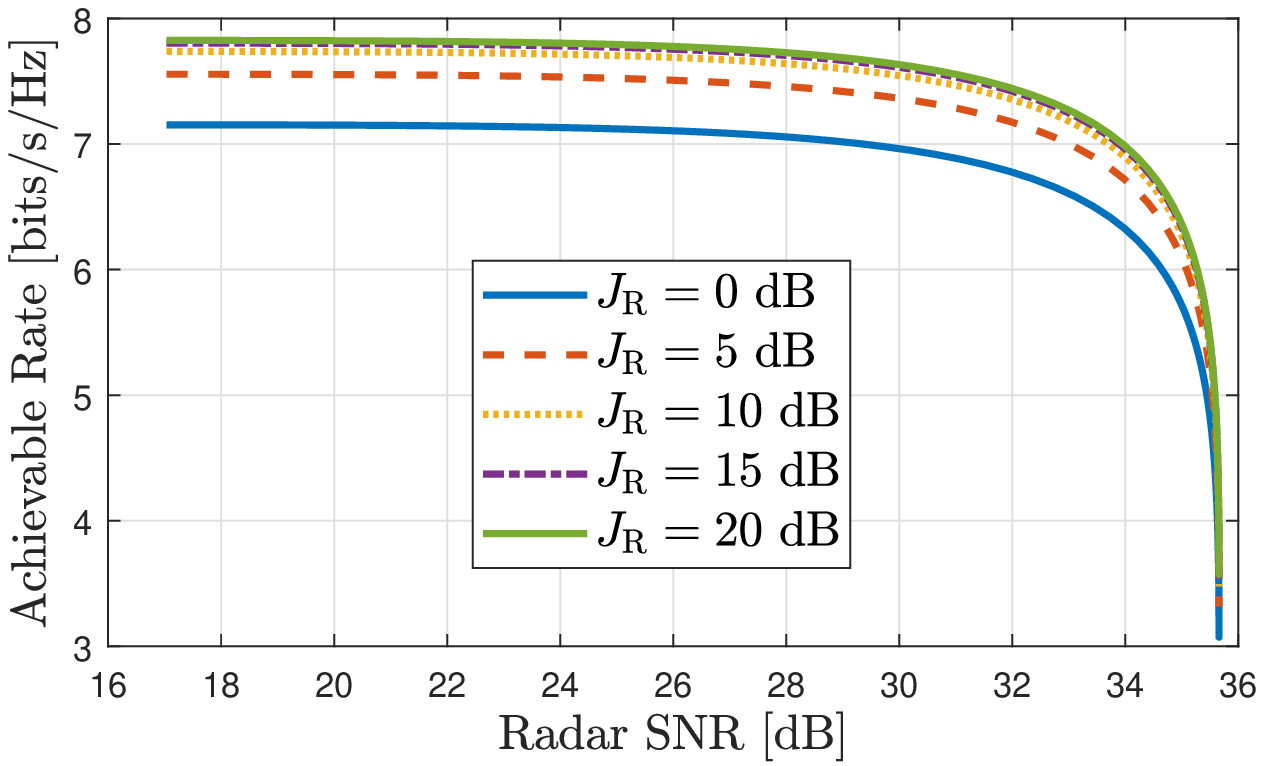} 
        \vspace{-0.2in}
        \caption{OTFS ISAC trade-off curves obtained by Alg.~\ref{alg_tradeoff} for different LMRs $\Kric$ as $\rho$ in \eqref{eq_problem_tradeoff2} changes over the interval $[0, 1]$.}
        \label{fig_tradeoff_curves}
    \end{minipage}
\end{figure}

In Fig.~\ref{fig_beampattern_tradeoff}, we illustrate the transmit beampatterns corresponding to the ISAC beamformer $\betabopt$ obtained at the output of Alg.~\ref{alg_tradeoff} for different trade-off values $\rho$. In compliance with the ISAC trade-off optimization in \eqref{eq_problem_tradeoff2}, the optimal beamformer dominantly illuminates the direction of the communication paths for small $\rho$, while the transmit power is focused more towards the direction of the radar targets as $\rho$ increases. It is seen that $\rho = 0.4$ yields a favorable trade-off between sensing and communications when the two functionalities are assigned similar importance levels.

To investigate ISAC trade-offs under different DD channel characteristics, we plot in Fig.~\ref{fig_tradeoff_curves} the ISAC trade-off curves\footnote{These curves represent the results of DD-domain water-filling in \eqref{eq_problem_tradeoff_subpp2} and are observed to coincide with those obtained through DD-domain uniform power allocation, i.e., $\qq = \boldone$. The reason is that $g_i$ in \eqref{eq_problem_tradeoff_subpp_obj2}, the diagonal values of the DD correlation matrix $\boldG$ in \eqref{eq_gmat}, are very close to each other, making the uniform allocation a near-optimal strategy for rate maximization. This is similar to the convergence of uniform power allocation to water-filling in frequency domain for OFDM at high SNRs \cite{waterFill_SNR_2011}.} obtained through Alg.~\ref{alg_tradeoff} for different $\Kric$ as $\rho$ varies over $[0,1]$. It is observed that the achievable rate improves with increasing $\Kric$, leading to  more favorable ISAC trade-offs for channels with larger $\Kric$ values\footnote{Notice that while varying $\Kric$, the total communication SNR $\sum_{k=0}^{\Ktilde-1} \abs{\alphat_k}^2 / \sigma^2$ is kept constant and all paths have the same angle. Hence, OTFS can exhibit different ISAC trade-off behavior for different $\Kric$ only due to changes in the DD domain channel properties.}. Hence, enhancing the sparsity of the DD domain communication channel (moving towards a more LOS-dominant channel as $\Kric$ becomes larger) increases the achievable rate, in compliance with the results in the OTFS literature (e.g., \cite{windowDesign_OTFS_TCOM_2021,otfs_comml_2023,OTFS_TVT_2023}). Intuitively, channel sparsity facilitates both channel estimation and data detection due to reduced channel spreading in delay and Doppler domains \cite{windowDesign_OTFS_TCOM_2021}. A rigorous explanation of this phenomenon follows from \eqref{eq_boldht_cov}: As the channel becomes more LOS-dominant, the cross-correlation terms in \eqref{eq_boldht_cov} vanish, making $\boldHt^H \boldHt $ a scaled identity matrix. This implies that $\boldG$ in \eqref{eq_gmat} becomes a scaled identity matrix, as well, making $(\rlmmmse(\pp, \betab))^{-1}$ in \eqref{eq_rlmmse_lem} more diagonally dominant, which in turn decreases $\log \det \rlmmmse(\pp, \betab)$ and increases the achievable rate in \eqref{eq_cap_lmmse}.



\section{Concluding Remarks}\vspace{-0.05in}
In this paper, we have performed an in-depth investigation of MIMO-OTFS ISAC systems, and proposed novel signal models, radar sensing and signal design algorithms, and ISAC transmission strategies. First, novel radar and communication signal models for OTFS have been derived by explicitly taking into account the impact of ISI and ICI, which unveils valuable analytical insights into how these effects manifest themselves in an OTFS system. Based on the new OTFS model, we have developed a GLRT-based 3-D sensing algorithm that exploits ISI and ICI to enhance delay-Doppler estimation performance. Moreover, we have proposed a novel strategy, called DD multiplexing, for the search mode of the proposed OTFS ISAC system, and an algorithm for ISAC signal design in spatial and DD domains for the track mode. In light of the simulation results, key insights from this work can be listed as follows: 
    \textit{(i) ISI/ICI Mitigation:} The proposed OTFS radar sensing algorithm can effectively mitigate the masking effect of ISI/ICI on weak targets, and provides superior detection performance compared to standard FFT-based methods.  
    \textit{(ii) ISI/ICI Exploitation:} ISI and ICI effects in OTFS radar can be exploited to significantly extend the unambiguous intervals in range and velocity, respectively. This enables detection of far-away/high-speed targets beyond the conventional ambiguity limits dictated by subcarrier spacing \cite{RadCom_Proc_IEEE_2011,OFDM_Radar_Phd_2014,OTFS_RadCom_TWC_2020,beamspaceMIMO_OTFS_2022} and estimation of their true range/velocity.  
    \textit{(iii) DD Multiplexing for MIMO-OTFS:} The proposed DD multiplexing strategy provides orthogonal transmission from TX antennas, creating a virtual ULA to improve angular resolvability of targets in the search mode.
    \textit{(iv) Spatial Signal Design for MIMO-OTFS:} The developed design strategy for ISAC TX beamforming in the track mode can offer flexible trade-offs with regard to illumination of radar targets and communication channel paths.
    \textit{(v) DD Channel Sparsity vs. ISAC Trade-offs:} The sparser the communication channel in the DD domain, the more favorable the OTFS ISAC trade-off. This results from the fact that the achievable rate of OTFS increases with the relative power of the LOS path under the constraint on the total power of LOS and NLOS paths. 
Future research will focus on extending the proposed MIMO-OTFS ISAC methods to multi-user scenarios.

\begin{appendices}

\section{Proof of Lemma~\ref{lemma_dd}}\label{app_lemma_dd}
Using \eqref{eq_xi_dd} and \eqref{eq_sssi}, the cross-correlation between the signals transmitted by the $\thn{i}$ and the $\thn{j}$ antennas with the DD multiplexing \eqref{eq_dd_mult} is given by
\begin{align} \label{eq_xcorr_ant}
    \big\langle \sss_i , \sss_j  \big\rangle &\triangleq \sss_i^H \sss_j = \tracesmall{ \left[ \gtxmat (\boldXdd \odot \boldW_i) \FF_M^H \right]^H \left[ \gtxmat (\boldXdd \odot \boldW_j) \FF_M^H \right] } \\ \nonumber
    &=\tracebig{  \FF_M^H  \FF_M  (\boldXdd \odot \boldW_i)^H \gtxmat^H   \gtxmat (\boldXdd \odot \boldW_j)  } = \tracebig{    (\boldXdd \odot \boldW_i)^H  (\boldXdd \odot \boldW_j)  } \\ \nonumber
     &= \boldone^T \Big(\conj{(\boldXdd)} \odot \boldW_i \odot \boldXdd \odot \boldW_j \Big) \boldone = P_i \, \delta(i-j)  ~,
\end{align}
which establishes the result in \eqref{eq_shs}.

\section{Computation of Radar SNR}\label{app_snr}
The radar SNR in the presence of a single target can be defined, according to \eqref{eq_obs_mimo}, as
\begin{align} \nonumber
    \snrrad(\pp, \betab; \tau, \nu, \theta )    
    &= \frac{ \Eee\big\{ \norm{\alpha     \boldC(\nu)  \FF^H \boldB(\tau) \FF \boldS  \atx(\theta) \arx^T(\theta)}_F^2 \big\} }{ \Eee\big\{ \norm{\boldZ}_F^2 \big\} } ~,
    \\ \nonumber
    &= \frac{\abs{\alpha}^2}{\sigma^2 NM} \traceee \Big( \Eee\big\{  \arx^{\ast}(\theta) \atx^H(\theta) \boldS^H \FF^H \boldB^H(\tau)  \FF \boldC^H(\nu)   \boldC(\nu)  \FF^H \boldB(\tau) \FF \boldS  \atx(\theta) \arx^T(\theta) \big\} \Big) ~,
    \\ \nonumber
    &= \frac{\abs{\alpha}^2}{\sigma^2 NM} \traceee \Big( \norm{\arx(\theta)}^2 \Eee\big\{   \atx^H(\theta) \boldS^H  \boldS  \atx(\theta) \big\}   \Big) = \frac{\abs{\alpha}^2}{\sigma^2 NM} \Eee\big\{ \atx^H(\theta) \boldS^H  \boldS  \atx(\theta) \big\}   ~.
\end{align}
Using \eqref{eq_sat_der}, the SNR becomes
\begin{align} \nonumber
    \snrrad(\pp, \betab; \tau, \nu, \theta )
    &=  \frac{\abs{\alpha}^2}{\sigma^2 NM} \Eee\big\{ \Big[ (\FF_M^H \otimes \Imatrix_N) \betab^T \atx(\theta) \diag{\pp} \xxdd \Big]^H  
      \Big[ (\FF_M^H \otimes \Imatrix_N) \betab^T \atx(\theta) \diag{\pp} \xxdd \Big]  \big\} ~,
    \\ \nonumber
    &= \frac{\abs{\alpha}^2 \abs{ \betab^T \atx(\theta)}^2}{\sigma^2 NM} \Eee\big\{  (\xxdd)^H \diag{\pp \odot \pp} \xxdd  \big\} 
    = \frac{\abs{\alpha}^2 \abs{ \betab^T \atx(\theta)}^2}{\sigma^2 NM} \ppbar^T \Eee\big\{ \xxddbar \big\} ~,
    \\ \label{eq_snr_rad}
    &= \frac{\abs{\alpha}^2 \abs{ \betab^T \atx(\theta)}^2}{\sigma^2} ~,
\end{align}
where $\ppbar \triangleq \pp \odot \pp$ and $\xxddbar \triangleq \xxdd \odot (\xxdd)^{\ast}$ denote, respectively, the powers allocated to the DD bins and the powers of the DD data symbols. The last step in \eqref{eq_snr_rad} stems from $\Eee\big\{ \xxddbar \big\} = 1$ and $\ppbar^T \boldone = NM$.

\section{Proof of Lemma~\ref{lemma_lmmse}}\label{app_proof_lmmse}

With \eqref{eq_wwi_phased}, we have $ \boldW \atx(\thetat_k) = \sum_{i=1}^{\Ntx}  \ww_i [\atx(\thetat_k)]_i =  \betab^T \atx(\thetat_k) \pp$, 
which allows re-expressing \eqref{eq_hdd} as
\begin{align} \label{eq_hdd2}
    \boldHdd &= (\FF_M \otimes \Imatrix_N) \boldHt (\FF_M^H \otimes \Imatrix_N) \diagb{\pp } ~,
\end{align}
where $\boldHt$ is the time-domain channel matrix defined in \eqref{eq_ht_mat}. Plugging \eqref{eq_hdd2} into \eqref{eq_lmmse_cov}, the LMMSE covariance matrix can be written as a function of $\pp$ and $\betab$ as follows:
\begin{align} \nonumber
    \rlmmmse(\pp, \betab) 
    &= \Big( \Imatrix + \frac{1}{\sigma^2}  \diagb{\pp }^H (\FF_M \otimes \Imatrix_N)  \boldHt^H
   (\FF_M^H \otimes \Imatrix_N) (\FF_M \otimes \Imatrix_N) \boldHt (\FF_M^H \otimes \Imatrix_N) \diagb{\pp } \Big)^{-1} ~,
    \\ \label{eq_rlmmse_p_beta}
    &= \Big( \Imatrix + \frac{1}{\sigma^2}  \diagb{\pp } (\FF_M \otimes \Imatrix_N)  \boldHt^H
      \boldHt (\FF_M^H \otimes \Imatrix_N) \diagb{\pp } \Big)^{-1} 
    = \big( \Imatrix + (\pp \pp^T) \odot \boldG \big)^{-1} ~,
\end{align}
where $\boldG$ is as defined in \eqref{eq_gmat}.

\section{Proof of Lemma~\ref{lemma_approx_beta}}\label{app_diag}
The covariance matrix of $\boldHt$ can be obtained from \eqref{eq_ht_mat} as
\begin{align}\label{eq_boldht_cov}
    \boldHt^H \boldHt &= \sum_{k=0}^{\Ktilde-1}  \abs{\alphat_k}^2  \abs{\betab^T \atx(\thetat_k)}^2 \Imatrix_{NM}
    \\ \nonumber
    &+ \sum_{k_1=0}^{\Ktilde-1} \sum_{\substack{k_2=0 \\ k_2 \neq k_1}}^{\Ktilde-1} \Big[ \conj{\alphat}_{k_2} \alphat_{k_1}  \betab^T \atx(\thetat_{k_1})  \atx^H(\thetat_{k_2}) \conj{\betab}
 \FF^H \boldB^H(\taut_{k_2}) \FF \boldC(\nut_{k_1} - \nut_{k_2}) \FF^H \boldB(\taut_{k_1}) \FF \Big] ~.
\end{align}
Since $\boldB(\tau)$ is diagonal, the matrix $\FF^H \boldB(\tau) \FF$ is circulant and represents the channel matrix constructed from the time-domain channel impulse response $\FF^H \bb(\tau)$ corresponding to $\tau$. For small $M$, large $\deltaf$ or small Doppler spread, the Doppler shifts in $\boldC(\nut_{k_1} - \nut_{k_2})$ will be small, leading to the approximation $\boldC(\nut_{k_1} - \nut_{k_2}) \approx \Imatrix_{NM}$. Then, the matrix on the third line of \eqref{eq_boldht_cov} becomes approximately $\FF^H \boldB^H(\taut_{k_2}) \FF  \FF^H \boldB(\taut_{k_1}) \FF = \FF^H \boldB(\deltatau) \FF$, which is a circulant time-domain channel matrix corresponding to the delay $\deltatau \triangleq \taut_{k_1}-\taut_{k_2}$, as noted above. Hence, for $\deltatau \neq 0$, the principal diagonal of $\FF^H \boldB(\Delta_{\tau}) \FF$ will have small magnitude, while only the upper/lower diagonal corresponding to $\deltatau$ will have large magnitude (however, being a cross-term, its magnitude is still smaller than that of the diagonals on the first line of \eqref{eq_boldht_cov}). This implies that the first (direct) term in \eqref{eq_boldht_cov} adds up the contributions from all the paths on its diagonal, whereas the second one (cross-term) spreads the energy of the path cross-correlations across the different off-diagonals, leading to an approximately diagonal matrix \cite{zohair_TWC_2018} $\boldHt^H \boldHt \approx \sum_{k=0}^{\Ktilde-1}  \abs{\alphat_k}^2  \abs{\betab^T \atx(\thetat_k)}^2 \Imatrix_{NM}$. 
Plugging this into \eqref{eq_gmat} and \eqref{eq_rlmmse_lem}, we obtain
\begin{align} \nonumber
    \log \det & \rlmmmse(\pp, \betab) 
    \approx \log \det \Big( \Imatrix + \frac{1}{\sigma^2} \diagb{\pp} (\FF_M \otimes \Imatrix_N) \sum_{k=0}^{\Ktilde-1}  \abs{\alphat_k}^2  \abs{\betab^T \atx(\thetat_k)}^2
     (\FF_M^H \otimes \Imatrix_N) \diagb{\pp }   \Big)^{-1} \,,
    \\ \nonumber
    &= \log \det \Big( \Imatrix + \frac{1}{\sigma^2} \diagb{\qq} \sum_{k=0}^{\Ktilde-1}  \abs{\alphat_k}^2  \abs{\betab^T \atx(\thetat_k)}^2       \Big)^{-1} = - \sum_{i=0}^{NM-1} \log\big( 1 + q_i \betab^T \DDcom \conj{\betab} \big) \,.
\end{align}

\section{Proof of Lemma~\ref{lemma_lb}}\label{app_proof_lb}

Assuming that $\boldG$ has small off-diagonal elements, we can decompose it as \cite{sensorCorr_Varshney_2016}
\begin{align} \label{eq_G_dec}
    \boldG = \boldLambda + \epsilon \boldUps ~,
\end{align}
where $\boldLambda = \diag{\ggb}$ with $\ggb \in \realset{NM}{1}$ denoting the diagonal elements of $\boldG$ and $\epsilon \boldUps$ is a hollow matrix \cite{zohair_TWC_2018} whose diagonal elements are zero and off-diagonal ones equal to those of $\boldG$. Here, $\epsilon$ has a small value. Using the representation in \eqref{eq_G_dec}, the achievable rate in \eqref{eq_rlmmse_lem} can be expressed as
\begin{align} \label{eq_logdet}
    -\log \det \rlmmmse(\pp) 
    = \log \det \big( \Imatrix + (\pp \pp^T) \odot (\boldLambda + \epsilon \boldUps) \big) 
    ~= \log \det \big( \Imatrix + \diag{\qq \odot \ggb} + \epsilon (\pp \pp^T) \odot  \boldUps \big) ~. 
\end{align}
In \eqref{eq_logdet}, using the first-order approximation of $\log \det (\Imatrix + \boldX)$ around $\boldX_0$ with $\boldX = \boldX_0 + \epsilon (\pp \pp^T) \odot  \boldUps$ and $\boldX_0 =  \Imatrix + \diag{\qq \odot \lambdab}$ yields \cite[Eq.~(17)]{DFRC_BF_JSAC_2022}
\begin{align} \label{eq_logdet2}
    &-\log \det \rlmmmse(\pp) \approx \log \det \big(\Imatrix + \diag{\qq \odot \ggb}\big) 
    + \tracebig{ \big[\Imatrix + \diag{\qq \odot \ggb}\big]^{-1}    \big[\epsilon (\pp \pp^T) \odot  \boldUps \big] } ~.
\end{align}
Due to $\boldUps$ being a hollow matrix, the second term on the right-hand side of \eqref{eq_logdet2} is zero. Hence, the achievable rate can be approximated as $-\log \det \rlmmmse(\pp, \betab) \approx \log \det \big(\Imatrix + \diag{\qq \odot \ggb}\big)$. 

\end{appendices}

\bibliographystyle{IEEEtran}
\bibliography{otfs}

\begin{thebibliography}{10}
\providecommand{\url}[1]{#1}
\csname url@samestyle\endcsname
\providecommand{\newblock}{\relax}
\providecommand{\bibinfo}[2]{#2}
\providecommand{\BIBentrySTDinterwordspacing}{\spaceskip=0pt\relax}
\providecommand{\BIBentryALTinterwordstretchfactor}{4}
\providecommand{\BIBentryALTinterwordspacing}{\spaceskip=\fontdimen2\font plus
\BIBentryALTinterwordstretchfactor\fontdimen3\font minus
  \fontdimen4\font\relax}
\providecommand{\BIBforeignlanguage}[2]{{%
\expandafter\ifx\csname l@#1\endcsname\relax
\typeout{** WARNING: IEEEtran.bst: No hyphenation pattern has been}%
\typeout{** loaded for the language `#1'. Using the pattern for}%
\typeout{** the default language instead.}%
\else
\language=\csname l@#1\endcsname
\fi
#2}}
\providecommand{\BIBdecl}{\relax}
\BIBdecl

\bibitem{6g_vision_2023}
C.-X. Wang \emph{et~al.}, ``On the road to {6G}: Visions, requirements, key
  technologies and testbeds,'' \emph{IEEE Communications Surveys \& Tutorials},
  pp. 1--1, 2023.

\bibitem{6g_hexax}
M.~A. Uusitalo \emph{et~al.}, ``{6G} vision, value, use cases and technologies
  from {European} {6G} flagship project {Hexa-X},'' \emph{IEEE Access}, vol.~9,
  pp. 160\,004--160\,020, 2021.

\bibitem{6g_wp3_hexax}
A.~Behravan \emph{et~al.}, ``Positioning and sensing in {6G}: Gaps, challenges,
  and opportunities,'' \emph{IEEE Vehicular Technology Magazine}, vol.~18,
  no.~1, pp. 40--48, 2023.

\bibitem{Fan_ISAC_6G_JSAC_2022}
F.~Liu \emph{et~al.}, ``Integrated sensing and communications: Toward
  dual-functional wireless networks for {6G} and beyond,'' \emph{IEEE Journal
  on Selected Areas in Communications}, vol.~40, no.~6, pp. 1728--1767, 2022.

\bibitem{Lima6Gsensing20}
C.~De~Lima \emph{et~al.}, ``Convergent communication, sensing and localization
  in {6G} systems: An overview of technologies, opportunities and challenges,''
  \emph{IEEE Access}, vol.~9, pp. 26\,902--26\,925, 2021.

\bibitem{wymeersch2020radio}
H.~Wymeersch \emph{et~al.}, ``Radio localization and mapping with
  reconfigurable intelligent surfaces: Challenges, opportunities, and research
  directions,'' \emph{IEEE Vehicular Technology Magazine}, vol.~15, no.~4, pp.
  52--61, 2020.

\bibitem{5g_6g_isac_2021}
T.~Wild \emph{et~al.}, ``Joint design of communication and sensing for beyond
  {5G} and {6G} systems,'' \emph{IEEE Access}, vol.~9, pp. 30\,845--30\,857,
  2021.

\bibitem{banelli2014modulation}
P.~Banelli \emph{et~al.}, ``Modulation formats and waveforms for {5G} networks:
  Who will be the heir of {OFDM}?: An overview of alternative modulation
  schemes for improved spectral efficiency,'' \emph{IEEE Signal Processing
  Magazine}, vol.~31, no.~6, pp. 80--93, 2014.

\bibitem{ofdm_radar_correlation_TAES_2020}
S.~Mercier \emph{et~al.}, ``Comparison of correlation-based {OFDM} radar
  receivers,'' \emph{IEEE Transactions on Aerospace and Electronic Systems},
  vol.~56, no.~6, pp. 4796--4813, 2020.

\bibitem{MIMO_OFDM_ICI_JSTSP_2021}
M.~F. Keskin \emph{et~al.}, ``{MIMO-OFDM} joint radar-communications: Is {ICI}
  friend or foe?'' \emph{IEEE Journal of Selected Topics in Signal Processing},
  vol.~15, no.~6, pp. 1393--1408, 2021.

\bibitem{papr_otfs_2022}
P.~Wei \emph{et~al.}, ``Charactering the peak-to-average power ratio of {OTFS}
  signals: A large system analysis,'' \emph{IEEE Transactions on Wireless
  Communications}, vol.~21, no.~6, pp. 3705--3720, 2022.

\bibitem{hadani2017orthogonal}
R.~Hadani \emph{et~al.}, ``Orthogonal time frequency space modulation,'' in
  \emph{2017 IEEE Wireless Communications and Networking Conference
  (WCNC)}.\hskip 1em plus 0.5em minus 0.4em\relax IEEE, 2017, pp. 1--6.

\bibitem{otfs_ofdm_comp_TWC_2022}
L.~Gaudio \emph{et~al.}, ``{OTFS} vs. {OFDM} in the presence of sparsity: A
  fair comparison,'' \emph{IEEE Transactions on Wireless Communications},
  vol.~21, no.~6, pp. 4410--4423, 2022.

\bibitem{OTFS_SBL_TWC_2022}
Z.~Wei \emph{et~al.}, ``Off-grid channel estimation with sparse {Bayesian}
  learning for {OTFS} systems,'' \emph{IEEE Transactions on Wireless
  Communications}, pp. 1--1, 2022.

\bibitem{OTFS_CE_TSP_2019}
W.~Shen \emph{et~al.}, ``Channel estimation for orthogonal time frequency space
  ({OTFS}) massive {MIMO},'' \emph{IEEE Transactions on Signal Processing},
  vol.~67, no.~16, pp. 4204--4217, 2019.

\bibitem{isac_otfs_jstsp_2021}
W.~Yuan \emph{et~al.}, ``Integrated sensing and communication-assisted
  orthogonal time frequency space transmission for vehicular networks,''
  \emph{IEEE Journal of Selected Topics in Signal Processing}, vol.~15, no.~6,
  pp. 1515--1528, 2021.

\bibitem{OTFS_mag_2022}
S.~K. Mohammed \emph{et~al.}, ``{OTFS}—a mathematical foundation for
  communication and radar sensing in the delay-doppler domain,'' \emph{IEEE
  BITS the Information Theory Magazine}, vol.~2, no.~2, pp. 36--55, 2022.

\bibitem{lampel2022orthogonal}
F.~Lampel \emph{et~al.}, ``Orthogonal time frequency space modulation based on
  the discrete {Zak} transform,'' \emph{Entropy}, vol.~24, no.~12, p. 1704,
  2022.

\bibitem{ISAC_OTFS_JSAC_2022}
S.~Li \emph{et~al.}, ``A novel {ISAC} transmission framework based on
  spatially-spread orthogonal time frequency space modulation,'' \emph{IEEE
  Journal on Selected Areas in Communications}, vol.~40, no.~6, pp. 1854--1872,
  2022.

\bibitem{Gaudio_MIMO_OTFS_Hybrid}
L.~Gaudio \emph{et~al.}, ``Hybrid digital-analog beamforming and {MIMO} radar
  with {OTFS} modulation,'' \emph{arXiv preprint arXiv:2009.08785}, 2020.

\bibitem{otfs_radar_2019}
P.~Raviteja \emph{et~al.}, ``Orthogonal time frequency space {(OTFS)}
  modulation based radar system,'' in \emph{2019 IEEE Radar Conference
  (RadarConf)}.\hskip 1em plus 0.5em minus 0.4em\relax IEEE, 2019, pp. 1--6.

\bibitem{OTFS_RadCom_TWC_2020}
L.~Gaudio \emph{et~al.}, ``On the effectiveness of {OTFS} for joint radar
  parameter estimation and communication,'' \emph{IEEE Transactions on Wireless
  Communications}, vol.~19, no.~9, pp. 5951--5965, 2020.

\bibitem{MIMO_OTFS_Radar_2020}
------, ``Joint radar target detection and parameter estimation with {MIMO
  OTFS},'' in \emph{2020 IEEE Radar Conference (RadarConf20)}, 2020, pp. 1--6.

\bibitem{OTFS_IOT_2021}
K.~Wu \emph{et~al.}, ``{OTFS}-based joint communication and sensing for future
  industrial {IoT},'' \emph{IEEE Internet of Things Journal}, vol.~10, no.~3,
  pp. 1973--1989, 2023.

\bibitem{beamspaceMIMO_OTFS_2022}
S.~K. Dehkordi \emph{et~al.}, ``Beam-space {MIMO} radar for joint communication
  and sensing with {OTFS} modulation,'' \emph{IEEE Transactions on Wireless
  Communications}, pp. 1--1, 2023.

\bibitem{OTFS_ISAC_part3_2022}
W.~Yuan \emph{et~al.}, ``Orthogonal time frequency space modulation—part iii:
  {ISAC} and potential applications,'' \emph{IEEE Communications Letters},
  vol.~27, no.~1, pp. 14--18, 2023.

\bibitem{ofdm_otfs_comparison_2022}
A.~Correas-Serrano \emph{et~al.}, ``Comparison of radar receivers for {OFDM}
  and {OTFS} waveforms,'' in \emph{2022 19th European Radar Conference
  (EuRAD)}, 2022, pp. 1--4.

\bibitem{OTFS_Eq_Learning_TWC_2022}
Z.~Zhou \emph{et~al.}, ``Learning to equalize {OTFS},'' \emph{IEEE Transactions
  on Wireless Communications}, vol.~21, no.~9, pp. 7723--7736, 2022.

\bibitem{OTFS_ICC_Workshop_2021}
M.~F. Keskin \emph{et~al.}, ``Radar sensing with {OTFS}: Embracing {ISI} and
  {ICI} to surpass the ambiguity barrier,'' in \emph{2021 IEEE International
  Conference on Communications Workshops (ICC Workshops)}, 2021, pp. 1--6.

\bibitem{jointRadCom_review_TCOM}
F.~{Liu} \emph{et~al.}, ``Joint radar and communication design: Applications,
  state-of-the-art, and the road ahead,'' \emph{IEEE Transactions on
  Communications}, vol.~68, no.~6, pp. 3834--3862, 2020.

\bibitem{reducedCP_OTFS_2018}
P.~Raviteja \emph{et~al.}, ``Practical pulse-shaping waveforms for
  reduced-cyclic-prefix {OTFS},'' \emph{IEEE Transactions on Vehicular
  Technology}, vol.~68, no.~1, pp. 957--961, 2018.

\bibitem{otfs_frac_2020}
Y.~Ge \emph{et~al.}, ``Receiver design for {OTFS} with a fractionally spaced
  sampling approach,'' \emph{IEEE Transactions on Wireless Communications},
  vol.~20, no.~7, pp. 4072--4086, 2021.

\bibitem{sensingAssistedComm_TWC_2020}
F.~Liu \emph{et~al.}, ``Radar-assisted predictive beamforming for vehicular
  links: Communication served by sensing,'' \emph{IEEE Transactions on Wireless
  Communications}, vol.~19, no.~11, pp. 7704--7719, 2020.

\bibitem{ISAC_V2I_Extended_TWC_2022}
Z.~Du \emph{et~al.}, ``Integrated sensing and communications for {V2I}
  networks: Dynamic predictive beamforming for extended vehicle targets,''
  \emph{IEEE Transactions on Wireless Communications}, pp. 1--1, 2022.

\bibitem{multibeam_TVT_2019}
J.~A. Zhang \emph{et~al.}, ``Multibeam for joint communication and radar
  sensing using steerable analog antenna arrays,'' \emph{IEEE Transactions on
  Vehicular Technology}, vol.~68, no.~1, pp. 671--685, 2019.

\bibitem{DFRC_Waveform_Design}
F.~{Liu} \emph{et~al.}, ``Toward dual-functional radar-communication systems:
  Optimal waveform design,'' \emph{IEEE Transactions on Signal Processing},
  vol.~66, no.~16, pp. 4264--4279, Aug 2018.

\bibitem{JCR_JSTSP_2021}
P.~Kumari \emph{et~al.}, ``Adaptive and fast combined waveform-beamforming
  design for mmwave automotive joint communication-radar,'' \emph{IEEE Journal
  of Selected Topics in Signal Processing}, vol.~15, no.~4, pp. 996--1012,
  2021.

\bibitem{beamformer_ISAC_FD_TCOM_2022}
C.~B. Barneto \emph{et~al.}, ``Beamformer design and optimization for joint
  communication and full-duplex sensing at mm-waves,'' \emph{IEEE Transactions
  on Communications}, vol.~70, no.~12, pp. 8298--8312, 2022.

\bibitem{phasedMIMO_radar_TSP_2010}
A.~Hassanien \emph{et~al.}, ``Phased-{MIMO} radar: A tradeoff between
  phased-array and {MIMO} radars,'' \emph{IEEE Transactions on Signal
  Processing}, vol.~58, no.~6, pp. 3137--3151, 2010.

\bibitem{OFDM_OTFS_modem_2017}
A.~Farhang \emph{et~al.}, ``Low complexity modem structure for {OFDM}-based
  orthogonal time frequency space modulation,'' \emph{IEEE Wireless
  Communications Letters}, vol.~7, no.~3, pp. 344--347, 2017.

\bibitem{OTFS_Canc_Iterative_TWC_2018}
P.~Raviteja \emph{et~al.}, ``Interference cancellation and iterative detection
  for orthogonal time frequency space modulation,'' \emph{IEEE Transactions on
  Wireless Communications}, vol.~17, no.~10, pp. 6501--6515, 2018.

\bibitem{windowDesign_OTFS_TCOM_2021}
Z.~Wei \emph{et~al.}, ``Transmitter and receiver window designs for orthogonal
  time-frequency space modulation,'' \emph{IEEE Transactions on
  Communications}, vol.~69, no.~4, pp. 2207--2223, 2021.

\bibitem{80211_Radar_TVT_2018}
P.~{Kumari} \emph{et~al.}, ``{IEEE} 802.11ad-based radar: An approach to joint
  vehicular communication-radar system,'' \emph{IEEE Transactions on Vehicular
  Technology}, vol.~67, no.~4, pp. 3012--3027, April 2018.

\bibitem{Firat_OFDM_2012}
R.~F. {Tigrek} \emph{et~al.}, ``{OFDM} signals as the radar waveform to solve
  {Doppler} ambiguity,'' \emph{IEEE Transactions on Aerospace and Electronic
  Systems}, vol.~48, no.~1, pp. 130--143, Jan 2012.

\bibitem{OFDM_Radar_Phd_2014}
M.~Braun, ``{OFDM} radar algorithms in mobile communication networks,''
  \emph{Karlsruher Institutes f{\"u}r Technologie}, 2014.

\bibitem{SPM_JRC_2019}
K.~V. {Mishra} \emph{et~al.}, ``Toward millimeter-wave joint radar
  communications: A signal processing perspective,'' \emph{IEEE Signal
  Processing Magazine}, vol.~36, no.~5, pp. 100--114, Sep. 2019.

\bibitem{raviteja2018practical}
P.~Raviteja \emph{et~al.}, ``Practical pulse-shaping waveforms for
  reduced-cyclic-prefix {OTFS},'' \emph{IEEE Transactions on Vehicular
  Technology}, vol.~68, no.~1, pp. 957--961, 2018.

\bibitem{Passive_OFDM_2010}
C.~R. Berger \emph{et~al.}, ``Signal processing for passive radar using {OFDM}
  waveforms,'' \emph{IEEE Journal of Selected Topics in Signal Processing},
  vol.~4, no.~1, pp. 226--238, 2010.

\bibitem{OFDM_Passive_Res_2017_TSP}
L.~{Zheng} \emph{et~al.}, ``Super-resolution delay-{Doppler} estimation for
  {OFDM} passive radar,'' \emph{IEEE Transactions on Signal Processing},
  vol.~65, no.~9, pp. 2197--2210, May 2017.

\bibitem{RadCom_Proc_IEEE_2011}
C.~{Sturm} \emph{et~al.}, ``Waveform design and signal processing aspects for
  fusion of wireless communications and radar sensing,'' \emph{Proceedings of
  the IEEE}, vol.~99, no.~7, pp. 1236--1259, July 2011.

\bibitem{OFDM_DFRC_TSP_2021}
M.~F. Keskin \emph{et~al.}, ``Limited feedforward waveform design for {OFDM}
  dual-functional radar-communications,'' \emph{IEEE Transactions on Signal
  Processing}, vol.~69, pp. 2955--2970, 2021.

\bibitem{Visa_CFO_TSP_2006}
T.~Roman \emph{et~al.}, ``Blind frequency synchronization in {OFDM} via
  diagonality criterion,'' \emph{IEEE Transactions on Signal Processing},
  vol.~54, no.~8, pp. 3125--3135, 2006.

\bibitem{otfs_modem_2018}
A.~Farhang \emph{et~al.}, ``Low complexity modem structure for {OFDM}-based
  orthogonal time frequency space modulation,'' \emph{IEEE Wireless
  Communications Letters}, vol.~7, no.~3, pp. 344--347, 2017.

\bibitem{MIMO_OFDM_radar_TAES_2020}
G.~Hakobyan \emph{et~al.}, ``{OFDM-MIMO} radar with optimized nonequidistant
  subcarrier interleaving,'' \emph{IEEE Transactions on Aerospace and
  Electronic Systems}, vol.~56, no.~1, pp. 572--584, 2020.

\bibitem{comb_MIMO_OFDM_radar_2020}
B.~Nuss \emph{et~al.}, ``Frequency comb {MIMO OFDM} radar with nonequidistant
  subcarrier interleaving,'' \emph{IEEE Microwave and Wireless Components
  Letters}, vol.~30, no.~12, pp. 1209--1212, 2020.

\bibitem{richards2005fundamentals}
M.~A. Richards, \emph{Fundamentals of Radar Signal Processing}.\hskip 1em plus
  0.5em minus 0.4em\relax Tata McGraw-Hill Education, 2005.

\bibitem{OMP_mmWave_2016}
J.~Lee \emph{et~al.}, ``Channel estimation via orthogonal matching pursuit for
  hybrid {MIMO} systems in millimeter wave communications,'' \emph{IEEE
  Transactions on Communications}, vol.~64, no.~6, pp. 2370--2386, 2016.

\bibitem{LMMSE_capacity}
P.~{Stoica} \emph{et~al.}, ``On {MIMO} channel capacity: an intuitive
  discussion,'' \emph{IEEE Signal Processing Magazine}, vol.~22, no.~3, pp.
  83--84, May 2005.

\bibitem{LMMSE_TSP_2017}
J.~P. {González-Coma} \emph{et~al.}, ``{MSE} balancing in the {MIMO BC}:
  Unequal targets and probabilistic interference constraints,'' \emph{IEEE
  Transactions on Signal Processing}, vol.~65, no.~12, pp. 3293--3305, June
  2017.

\bibitem{imperfectCSI_MIMO_2010}
M.~Ding \emph{et~al.}, ``Maximum mutual information design for {MIMO} systems
  with imperfect channel knowledge,'' \emph{IEEE transactions on information
  theory}, vol.~56, no.~10, pp. 4793--4801, 2010.

\bibitem{goldsmith2005wireless}
A.~Goldsmith, \emph{Wireless Communications}.\hskip 1em plus 0.5em minus
  0.4em\relax Cambridge University Press, 2005.

\bibitem{RIS_Beamform_JSTSP_2022}
A.~Fascista \emph{et~al.}, ``{RIS}-aided joint localization and synchronization
  with a single-antenna receiver: Beamforming design and low-complexity
  estimation,'' \emph{IEEE Journal of Selected Topics in Signal Processing},
  vol.~16, no.~5, pp. 1141--1156, 2022.

\bibitem{waterFill_SNR_2011}
H.~{Moon}, ``Waterfilling power allocation at high {SNR} regimes,'' \emph{IEEE
  Transactions on Communications}, vol.~59, no.~3, pp. 708--715, 2011.

\bibitem{otfs_comml_2023}
S.~Li \emph{et~al.}, ``Orthogonal time frequency space modulation—part ii:
  Transceiver designs,'' \emph{IEEE Communications Letters}, vol.~27, no.~1,
  pp. 9--13, 2023.

\bibitem{OTFS_TVT_2023}
A.~Tusha \emph{et~al.}, ``Low complex inter-doppler interference mitigation for
  {OTFS} systems via global receiver windowing,'' \emph{IEEE Transactions on
  Vehicular Technology}, pp. 1--14, 2023.

\bibitem{zohair_TWC_2018}
Z.~Abu-Shaban \emph{et~al.}, ``Error bounds for uplink and downlink {3D}
  localization in {5G} millimeter wave systems,'' \emph{IEEE Transactions on
  Wireless Communications}, vol.~17, no.~8, pp. 4939--4954, 2018.

\bibitem{sensorCorr_Varshney_2016}
S.~Liu \emph{et~al.}, ``Sensor selection for estimation with correlated
  measurement noise,'' \emph{IEEE Transactions on Signal Processing}, vol.~64,
  no.~13, pp. 3509--3522, 2016.

\bibitem{DFRC_BF_JSAC_2022}
L.~Chen \emph{et~al.}, ``Generalized transceiver beamforming for {DFRC} with
  {MIMO} radar and {MU-MIMO} communication,'' \emph{IEEE Journal on Selected
  Areas in Communications}, vol.~40, no.~6, pp. 1795--1808, 2022.

\end{thebibliography}

\vspace{-0.1in}

\end{document}